\documentclass[showpacs]{revtex4}

\usepackage{graphics}
\usepackage{graphicx}

\def\m@thcombine#1#2{%
  \setbox0=\hbox{$#1$}
  \setbox1=\hbox{$#2$}
  \ifdim\wd0>\wd1
    \setbox0=\hbox to\wd1{\hss\box0\hss}
  \else
    \setbox1=\hbox to\wd0{\hss\box1\hss}
  \fi
  \mathop{\vcenter{
    \offinterlineskip\box0\box1}}}
\def\lesim{\m@thcombine<\sim}
\def\gesim{\m@thcombine>\sim}

\newcommand{\vecr}{\mbox{\boldmath $r$}}

\newcommand{\veck}{\mbox{\boldmath $k$}}
\newcommand{\vecp}{\mbox{\boldmath $p$}}

\newcommand{\del}{\partial}

\begin{document}

\title{
Spatial structure of neutron Cooper pair in low density uniform matter
}

\author{Masayuki Matsuo}

\affiliation{
Department of Physics, Faculty of Science, Niigata University,
Niigata 950-2181, Japan }

\date{\today}

\begin{abstract}
We analyze spatial structure of
the neutron Cooper pair in superfluid low-density uniform matter 
by means of BCS calculations employing 
a bare force and the effective Gogny interaction.  
It is shown that the Cooper pair exhibits
a strong spatial di-neutron correlation in a wide range of neutron density
$\rho/\rho_0\approx 10^{-4}-0.5$. 
This feature is related to the crossover behavior
between the pairing of the weak coupling BCS type and 
the Bose-Einstein condensation of bound neutron pairs. 
We also show that the zero-range delta interaction can describe the 
spatial structure of the neutron Cooper pair if the
density dependent interaction strength and the cut-off energy are appropriately chosen. 
Parameterizations of the density-dependent delta interaction 
satisfying this condition are discussed.
\end{abstract}


\maketitle

\section{Introduction}\label{Intro}

The importance of the pair correlation has been widely
recognized for nucleon many-body systems in various circumstances,
in particular in open-shell nuclei and in neutron stars. 
The pairing gap varies with 
system parameters such as $N$, $Z$ and the rotational frequency 
in the case of finite nuclei,
or the temperature and the density in the case of neutron stars
(cf. Refs.\cite{BM2,Shimizu89,TT93,Lombardo-Schulze,Dean03} as reviews). 
The pairing correlation at low nucleon density is of special interest since
the theoretical predictions for low-density uniform matter suggest
that the pairing gap may take, at around 1/10 of the normal nuclear
density, a value which is considerably
larger than that around the normal density
\cite{TT93,Lombardo-Schulze,Dean03}. 
This feature is expected to have direct relevance to the properties of neutron
stars, especially those associated with the inner crust\cite{Nstar1,Nstar2}.
The strong pairing at low density
may be relevant also to finite nuclei, if one considers
neutron-rich nuclei near the drip-line\cite{Bertsch91,DobHFB2,DD-Dob,DD-mix}. 
This is because such nuclei
often accompany  unsaturated low-density distribution of neutrons 
(the neutron skin and/or the neutron halo) surrounding the 
nuclear surface\cite{Tanihata,Tanihata-density,Ozawa}. 
It is interesting to clarify how 
the pair correlation in these exotic nuclei is different
from that in stable nuclei, reflecting the strong density dependence
mentioned above. In this connection we would like to ask
how the pair correlation at low nucleon density is different
from that around the normal density.

Spatial structure of the neutron Cooper pair is
focused upon as a characteristic feature of the low density nucleon pairing.
Its possible indication could be 
the di-neutron correlation in
the two-neutron halo nuclei, e.g. $^{11}$Li, for which
a spatially correlated pair formed by the halo neutrons
has been predicted theoretically
\cite{Bertsch91,Hansen,Ikeda,Zhukov,Barranco01,Aoyama,Hagino05}
and debated in experimental studies\cite{Sackett,Shimoura,Zinser,Ieki}.
Further, a recent theoretical analysis\cite{Matsuo05} using 
the Hartree-Fock-Bogoliubov (HFB) 
method\cite{Ring-Schuck,Blaizot-Ripka,DobHFB,Bulgac}
predicts also presence of  
similar di-neutron correlation in medium-mass neutron-rich nuclei
where more than two weakly-bound neutrons contribute to form 
the neutron skin 
in the exterior of the nuclear surface. 
It is also possible to argue importance
of the spatial correlation from a more fundamental viewpoint based 
on the nucleon 
interaction in the $^1S$ channel.
The bare nucleon-nucleon interaction in this 
channel has a virtual state around 
zero energy characterized by the large
scattering length $a\approx -18$ fm, which implies a very strong
attraction between a pair of neutrons with the spin singlet configurations. 
A rather general argument\cite{Leggett,Nozieres}, 
which applies to a dilute limit of 
any Fermion systems,
indicates that the pair correlation of the Fermions interacting with a large
scattering length differs largely from what is considered in the conventional
BCS theory\cite{BCS} assuming weak coupling: it is then appropriate
to consider a crossover 
between a superfluid system of the weak-coupling BCS type
and a Bose-Einstein condensate of spatially compact bound Fermion 
pairs\cite{Leggett,Nozieres,Melo,Engelbrecht,Randeria}. 
This BCS-BEC crossover phenomenon
was recently observed in a ultra-cold atomic gas in a trap
for which the interaction is controllable\cite{Regal}.
In the case of the nucleon pairing, 
the BCS-BEC crossover has been argued mostly
for the neutron-proton pairing in the $^{3}SD_{1}$ channel, 
for which the strong spatial correlation associated with the deuteron 
and the BEC of the deuterons 
may emerge\cite{Alm93,Stein95,Baldo95,Lombardo01a,Lombardo01b}.  
Concerning the neutron pairing in the 
$^{1}S$ channel, which we discuss in the present paper,
we may also expect that the strong coupling feature may lead to
the spatial di-neutron correlation
although the realization of the crossover could be marginal and depend
on the density\cite{Stein95}.

In the present paper, we would like to clarify how the
the spatial structure of the neutron Cooper pair varies with the 
density. For this purpose, we shall investigate the 
neutron pair correlation in symmetric nuclear matter and neutron matter
in the low density region. Uniform matter is of course a
simplification of
the actual nucleon configurations in finite nuclei and neutron stars.
However they have a great advantage as
one can solve the gap equation in this case without ambiguity for 
various interactions including the bare nucleon-nucleon forces 
with the repulsive core
\cite{TT93,Lombardo-Schulze,Dean03,Baldo95,Lombardo01a,Lombardo01b,
Takatsuka72,Takatsuka84,Baldo90,Oslo96,Khodel,DeBlasio97,Serra,Garrido01}
as well as effective interactions such as 
the Gogny force\cite{Serra,Kucharek89,Sedrakian03}, 
provided that the
BCS approximation (equivalent to the HFB in finite nuclei) is assumed.
It is straightforward then to
determine the wave function of the neutron Cooper pair 
from the solution of the gap equation
\cite{TT93,Baldo95,Lombardo01a,Takatsuka84,Baldo90,Oslo96,DeBlasio97,Serra}. 
This provides us with a good reference frame to study
the spatial structure of the Cooper pair as a function of the
density while
we do not intend to make precise predictions on other 
properties of neutron and symmetric nuclear matter.
We shall perform an analysis employing
both a bare force and the effective Gogny force\cite{Gogny},
and using a Hartree-Fock single-particle spectrum associated with the 
media. Our main conclusion will be that the spatial di-neutron correlation
is  strong 
in a wide range of the low density $\rho/\rho_0\approx 10^{-4}-0.5$,
independently on the adopted forces. 
We shall clarify the nature of the strong
spatial di-neutron correlation in terms of the BCS-BEC crossover
model.

We shall also examine a 
possibility of phenomenological description of the
spatially correlated neutron Cooper pair.  Here we consider
a contact force with a parametrized 
density dependent interaction strength, called often 
the density dependent delta interaction (DDDI)
\cite{Bertsch91,DobHFB2,DD-Dob,DD-mix,DDpair-Chas,DDpair-Tera,DDpair-Taj,Fayans96,Fayans00,Garrido}.
The parameters of the DDDI need to be determined from
some physical constraints. For example, the interaction strength has
been constrained by conditions to 
reproduce 
the experimentally extracted pairing gap 
in finite nuclei\cite{DobHFB2,DD-Dob,DD-mix,DDpair-Tera,DDpair-Taj,Fayans96,Fayans00}, or
the density-dependence of the neutron 
pairing gap in symmetric nuclear matter
and the $^1S$ 
scattering length \cite{Bertsch91,Garrido}.
It should be noted here that the contact force requires 
a cut-off energy, which needs to be treated as an additional model parameter.
Concerning the cut-off parameter, 
attentions have been paid in the previous studies 
to convergence properties of the
pairing correlation energy in finite nuclei\cite{DobHFB2}, 
to the energy dependence of the phase shift
\cite{Esbensen97}, or to the renormalization with respect to the
pairing gap\cite{Bulren,YuBulgac}.
In the present paper we shall take a different approach to the cut-off, i.e., 
we investigate relevance 
of the cut-off parameter to the spatial structure of the neutron Cooper pair.
It will be shown 
that the cut-off energy plays an important role
to describe the strong spatial correlation at low density.
Considering this as a physical 
constraint on the cut-off energy,
we shall derive 
new parameter sets of the DDDI.

Preliminary results of this work are reported in Ref.\cite{Matsupre}.

\section{Formulation}\label{Formulation-sec}

\subsection{BCS approximation}\label{gap-eq-sec}

We describe the neutron pair correlation in uniform neutron matter
and in uniform symmetric nuclear matter
by means of the BCS approximation, which is equivalent to the Bogoliubov's generalized
mean-field approach\cite{Ring-Schuck,Blaizot-Ripka}.
One of the basic equations is the gap equation, which is written in the momentum
representation as 
\begin{eqnarray}\label{gap-eq}
\Delta(p)&=& - {1 \over 2 (2\pi)^3}\int d \veck \tilde{v}(\vecp-\veck)
{\Delta(k) \over E(k)}, \\
E(k)&=&\sqrt{(e(k)-\mu)^2 + \Delta(k)^2}.
\end{eqnarray}
Here  
$\Delta(k)$ is the pairing gap dependent on the single-particle 
momentum $k$, while
$e(k)$ and $E(k)$ are the single-particle and the quasiparticle
energies. $\tilde{v}(\vecp-\veck)$ 
is the matrix element of the nucleon-nucleon interaction in the $^1S$ channel.
The gap equation needs to be solved together with the number equation 
\begin{equation}\label{num-eq}
\rho \equiv {k_F^3 \over 3\pi^2} = {1\over (2\pi)^3}  \int  d \veck 
\left(1 + {e(k)-\mu \over E(k)}\right), 
\end{equation}
which determines the relation between the neutron density $\rho$
(the Fermi momentum $k_F$)
and the chemical potential $\mu$. 
The solution of these equations defines the ground state
wave function of the BCS type 
and the static pairing properties at zero temperature 
for a given density $\rho$.

\begin{figure}[htbp]
\centerline{
\includegraphics[angle=270,width=8.5cm]{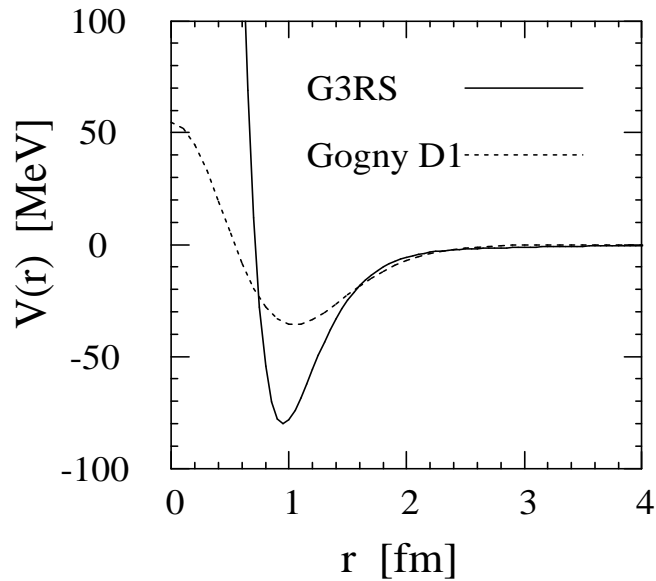}
}
\caption{
The nucleon-nucleon potential $v(r)$ in the ${}^1S$ channel
for the G3RS and Gogny D1 forces, plotted with the solid and
dotted curves, respectively, 
as a function of the relative distance $r$ between neutrons. 
\label{force}}
\end{figure}

As the interaction acting in the $^1S$ channel 
we shall adopt a bare nucleon-nucleon force,
the G3RS force\cite{Tamagaki}, 
and the effective interaction given by Gogny\cite{Gogny}.
The G3RS force is a local potential representation 
of the bare nucleon-nucleon interaction which is 
given by a superposition of three
Gaussian functions:
\begin{equation}\label{G3RS-eq}
v(\vecr) = \sum_i v_i e^{-r^2/\mu_i^2}.
\end{equation}
One component represents
a repulsive core with the height of 
$v_1=2000$ MeV and the range parameter $\mu_1=0.447$ fm, 
while two other Gaussians 
with $v_{2,3}=-240, -5$ MeV and  $\mu_{2,3}=0.942, 2.5$ fm 
represent the attraction dominant for $1\lesim r \lesim 3$ fm (see Fig.\ref{force}).
In spite of the simple three Gaussian representation, the G3RS reproduces
rather well the $^1S$ phase shift up to about $300$ MeV in the c.m. energy
of the scattering nucleons.
The associated scattering length
$a=-17.6$ fm is in close agreement with the experimental value
$a=-18.5\pm0.4$ fm\cite{Aexp}. The G3RS has been used in some 
BCS calculations for the $^{1}S$ pairing at low density
and for the $^{3}P_2$ pairing at high density\cite{TT93,Lombardo-Schulze,Dean03}. 
From a practical point of view, the analytic form makes it
easy to evaluate the
matrix elements of the interaction.

The Gogny force is an effective interaction
which is designed for the HFB description
of the pairing correlation in finite nuclei while keeping
some aspects of the $G$-matrix\cite{Gogny}. 
It is also a local potential represented
as a combination of two Gaussian functions in the form of 
Eq.(\ref{G3RS-eq}) with the
range parameters $\mu_{1,2}=0.7, 1.2$ fm.
In the following we show mostly results obtained with the 
parameter set D1\cite{Gogny} 
as we find no qualitative difference in the results for the 
parameter set D1S\cite{Gogny-D1S}. A common feature of the
G3RS and Gogny forces is that both are attractive 
in the range $1 \lesim r \lesim 3$ fm while they differ
largely for $r\lesim 0.5$ fm, where
the Gogny force exhibits only a 
very weak repulsion 
instead of the
short-range core present in the G3RS force (Fig.\ref{force}).
The interaction range of the two forces is 
of the order of 3fm. Note that
the experimental effective range is $r_{e}=2.80\pm0.11$ \cite{Aexp}.
We shall also apply a zero-range 
contact force $v(\vecr) \propto \delta(\vecr)$.
Our treatment of this interaction will be described separately in 
Section \ref{DDDI-sec}.

As the single-particle energy $e(k)$ we use, in the case of the Gogny interaction,
the Hartree-Fock single-particle
spectrum derived directly from the same interaction.
In the case of the bare force, it would be 
better  
from a viewpoint of the
self-consistency to use the Brueckner Hartree-Fock spectrum. 
But for simplicity 
we adopt in the present analysis an effective mass approximation.
Namely the single-particle energy is given by
$e(k)=k^2/2m^* $, where the effective mass 
$m^*=\left(\del^2 e(k) / \del^2 k|_{k_F}\right)^{-1}$ 
is derived from the 
Gogny HF spectrum\cite{Kucharek89} for the parameter set D1.

We solve the gap and number equations, (\ref{gap-eq}) and (\ref{num-eq}), 
without
introducing any cut-off. 
The momentum integrations in the two equations are performed
using a direct numerical method, where 
the maximum momentum $k_{max}$ for the integration is
chosen large enough so that the result does not depend on the 
choice of $k_{max}$. 
We adopt $k_{max}=20$ fm$^{-1}$
for the G3RS, and $k_{max}=10$ fm$^{-1}$ for the Gogny interaction.
Note here that it is dangerous to 
introduce a small energy window around 
the chemical potential (or the Fermi energy)  or a cut-off at a 
small momentum in evaluating the r.h.s. of
the gap equation.  Such an approximation may be justified only in the 
case of the weak coupling BCS  where the pairing gap is considerably 
smaller than the Fermi energy, but it is not applicable to the 
strong coupling case\cite{Leggett}.
Note also that 
the chemical potential $\mu$ 
and the Fermi energy $e_F$ are not the same except in the limit of the weak 
coupling. We define
the Fermi momentum $k_F$ through the 
nominal relation to the density $\rho =  {1 \over 3\pi^2} k_F^3$, 
and the Fermi energy $e_F$ by $e_F\equiv e(k_F)$.
The pairing gap 
$\Delta_F\equiv\Delta(k_F)$ at the Fermi momentum is used below
as a measure of the pair correlation.
In the following $\rho$ denotes always the neutron density.
(The total nucleon density is $\rho_{tot}=2\rho$ 
in the case of symmetric nuclear matter.)
We define, as a reference value, the normal neutron density by 
$\rho_0={1 \over 3\pi^2} k_{F0}^3$
with $k_{F0}=1.36$ fm$^{-1}$.

In order to investigate the spatial structure of the neutron Cooper
pair, it is useful to look into 
its wave function 
represented as a function of the relative distance between the partner neutrons of the pair.
It is given by 
\begin{eqnarray}\label{Cooper-eq}
\Psi_{pair}(r) &\equiv&
 C' \left<\Phi_0\right|\psi^\dagger(\vecr\uparrow)
\psi^\dagger(\vecr'\downarrow)\left|\Phi_0  \right>
= 
{C \over (2\pi)^3}
\int d \veck  u_kv_k e^{i\veck\cdot(\vecr-\vecr')}, \\
u_kv_k&=& {\Delta(k) \over 2E(k)},
\end{eqnarray}
in terms of the $u,v$-factors except by the normalization factors $C$ and $C'$. Here 
$\psi^\dagger(\vecr \sigma) \ \ (\sigma=\uparrow,\downarrow)$ is the creation operator of neutron and
$\left|\Phi_0\right>$ is the BCS ground state.
The Cooper pair wave function depends only on 
the relative distance $r=|\vecr-\vecr'|$ between the partners
as it is an s-wave. 
We evaluate the momentum integral in Eq.(\ref{Cooper-eq})
in the same way as in the gap and the number equations.

It is useful to evaluate the size of the neutron Cooper pair. 
A straightforward measure is the
r.m.s. radius of the Cooper pair
\begin{equation}\label{rms-eq}
\xi_{rms}= \sqrt{\left<r^2\right>}, 
\end{equation}
where
\begin{equation}\label{rms2-eq}
\left<r^2\right>= \int d\vecr r^2 |\Psi_{pair}(r)|^2 
= {\int_0^{\infty} dk k^2 \left({\del \over \del k}u_kv_k\right)^2 \over
\int_0^{\infty} dk k^2 \left(u_kv_k\right)^2}
\end{equation}
can be calculated directly from the Cooper pair wave function
$\Psi_{pair}(r)$ and/or from the $u,v$-factors in the momentum space. 
If one assumes weak coupling, the Pippard's coherence length \cite{BCS}
\begin{equation}\label{Pippard-eq}
\xi_P ={\hbar^2 k_F \over m^{*}\pi\Delta_F}
\end{equation}
given analytically in terms of the gap and the Fermi momentum may be
used also as another estimate of the size of the Cooper pair. In the following we mostly use
$\xi_{rms}$ since this quantity itself has a solid meaning even in the
case of the strong coupling BEC case and in the crossover region 
between BCS and BEC.
We shall use $\xi_{P}$ for qualitative discussions.

In the present paper, we neglect 
higher order many-body effects which go beyond
the BCS approximation. In many calculations
\cite{Chen86,Chen93,Ainsworth89,
Wambach93,Schulze96,Schulze01,Lombardo01,Shen03,Schwenk,Lombardo04} 
the higher order effects in low density neutron matter
are predicted to reduce the pairing gap by about a factor of two, which is
however very much dependent on the prescriptions 
adopted\cite{Lombardo-Schulze,Dean03} except
for the low density limit 
$\rho\rightarrow 0$\cite{Heiselberg00}. 
A recent Monte Carlo study\cite{Fabrocini05} 
using the realistic bare force suggests the gap close to the BCS result.
The higher order effects in symmetric nuclear matter\cite{Lombardo04}
and in finite nuclei\cite{Milan1,Milan2,Milan3}
are estimated to increase the gap. Keeping in
mind these ambiguities, we consider that 
the BCS approximation provides a meaningful
zero-th order reference.

\subsection{Pairing gap and coherence length}\label{gap-sec}

\begin{figure}[htbp]
\centerline{
\includegraphics[angle=270,width=8.5cm]{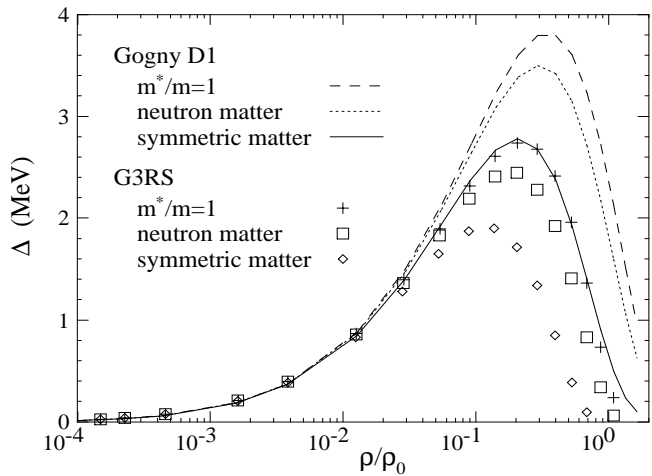}
}
\caption{
The pairing gap $\Delta_F$ in neutron and symmetric nuclear matter
as a function of the neutron density
$\rho/\rho_0$.
The results for the G3RS force are shown with the
symbols: cross for the free single-particle spectrum,
square for neutron matter, and diamond for
nuclear matter. The results with the Gogny D1 force
are plotted with the dashed, dotted, and 
solid curves for matter with the free single-particle spectrum, 
for neutron matter, and for symmetric matter, respectively.
\label{gap}}
\end{figure}

\begin{figure}[htbp]
\centerline{
\includegraphics[angle=270,width=8.5cm]{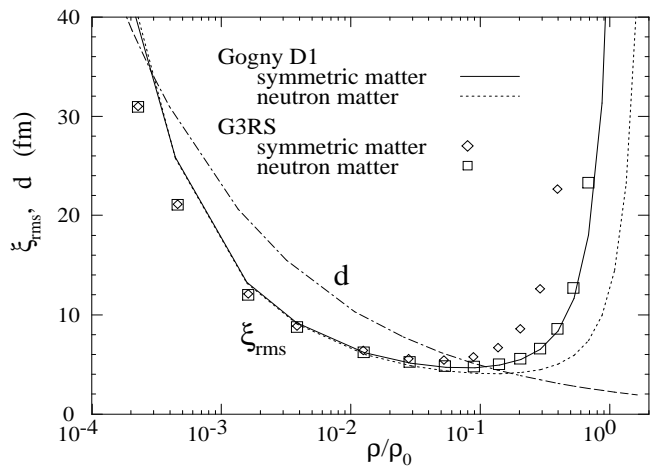}
}
\caption{
The r.m.s. radius $\xi_{rms}$ of the neutron Cooper pair
in uniform matter, plotted as a function of 
the neutron density $\rho/\rho_0$.
The results for symmetric nuclear and neutron matter
obtained with the Gogny D1 force are shown by the solid and dotted 
curves, respectively, while the results for symmetric nuclear and neutron
matter with the G3RS force are 
shown by the diamond and square symbols, respectively. 
The average inter-neutron
distance $d=\rho^{-1/3}$ is plotted with the dot-dashed line.
\label{rms}}
\end{figure}

Figure \ref{gap} shows 
the neutron pairing gap $\Delta_F$ 
obtained with the G3RS and Gogny D1 forces both 
for neutron matter and for symmetric matter.  Results
assuming the free single-particle spectrum (equivalent to the use
of $m^*=m$) are also shown for comparison. The pairing gap
becomes maximum around $\rho/\rho_0\sim 0.1-0.3$ in all the cases.
The gap decreases gradually
with further decreasing the density. The difference between 
neutron matter and symmetric nuclear matter, which originates from
the effective mass effect, becomes negligible at 
low density $\rho/\rho_0 \lesim 5\times 10^{-2}$.

The pairing gap obtained with the G3RS force is 
very similar to those obtained with more realistic models of
the bare force
(OPEG\cite{Takatsuka72,Takatsuka84,TT93}, 
Reid\cite{Takatsuka84,TT93,Khodel}, Argonne\cite{Khodel,Baldo90}, Paris\cite{Baldo90},
Bonn\cite{Oslo96,Serra}, Nijmegen\cite{Lombardo-Schulze}). 
This is because the gap is essentially determined by the $^{1}S$ phase shift
function\cite{phase-shift}, and the G3RS force reproduces the experimental phase shift,
though not as accurately as the modern forces.
There is small difference from them: the maximum gap 
$\Delta_F \approx 2.5$ MeV around $\rho/\rho_0 \sim 0.2$ 
(or $k_F \sim 0.8 {\rm fm}^{-1}$) for neutron matter is  
slightly smaller in the G3RS by about 5-20\%. 
The difference may be due to a limitation 
of the simple three-Gaussian representation, but
the quality is enough for the following discussions.
The gap obtained with the Gogny force is consistent
with those in the previous calculations\cite{Kucharek89,Serra,Sedrakian03}.
If we compare with the Gogny and G3RS results, they exhibit a
similar overall
density dependence, and 
a significant difference between the two forces is seen 
only at modest density
$\rho/\rho_0 \gesim 5\times 10^{-2}$. Garrido {\it et al.} 
\cite{Garrido} suggests that the similarity may indicate a possible
cancellation among higher order effects in the case of symmetric matter.
We find a less significant difference in the gap
at 
rather low density $\rho/\rho_0 \lesim 10^{-3}$, 
though not visible in Fig.\ref{gap}.
This arises from the 
fact that the scattering length $a=-13.5$ fm of the Gogny D1 
force deviates from the G3RS value $a=-17.6$ fm.

The r.m.s. radius $\xi_{rms}$ of the neutron Cooper pair 
calculated for neutron and symmetric nuclear matter with the 
bare or the Gogny forces are shown in Fig.\ref{rms} and Table \ref{coherence-length}.
The calculated $\xi_{rms}$ is consistent with the r.m.s. radius (or the Pippard's
coherence length) reported in the 
BCS calculations using other models of 
the bare force\cite{Takatsuka84,TT93,DeBlasio97,Serra}. Here we would like
emphasize characteristic density dependence of $\xi_{rms}$. It is seen
from Fig.\ref{rms} that 
$\xi_{rms}$ decreases
dramatically by nearly a factor of ten
from a large value
of the order of $\sim 50$ fm around the normal density $\rho/\rho_0\sim 1$
to considerably smaller values $\xi_{rms} \approx 4.5-6$ fm 
at density around
$\rho/\rho_0 \sim 0.1$. 
The size of the Cooper pair stays at small values 
$\xi_{rms} \approx 5-6$ fm in the density region $\rho/\rho_0 \sim 
10^{-2}-0.1$. It then turns to increase, but only gradually, at
further low density. These features are commonly seen
for both neutron and symmetric nuclear
matter, and for both the G3RS and Gogny forces.

We would like to emphasize also the smallness of the neutron Cooper pair. 
This may be elucidated if we 
compare the r.m.s. radius $\xi_{rms}$ 
with the average inter-neutron distance $d\equiv\rho^{-1/3}=3.09k_F^{-1}$.
It is seen from Fig.\ref{rms} that $\xi_{rms}$
becomes smaller than $d$ in a very wide range of density
$\rho/\rho_0 \sim 10^{-4} - 0.1$. 
The ratio $\xi_{rms}/d$ can reach values as small as
$\approx 0.5$  at density around $\rho/\rho_0 \sim 10^{-2}$.
The relation $\xi_{rms} <d$, i.e. 
the size of the neutron Cooper pair smaller than
the average inter-neutron distance, suggests that the neutron
Cooper pair exhibits strong spatial di-neutron correlation.

To understand this strong spatial correlation, 
it is useful to consider the ratio $\Delta_F/e_F$
between the pairing gap and the Fermi energy rather than the
absolute magnitude of the gap. The pairing gap 
$\Delta_F \approx 0.2$ MeV at the density $\rho/\rho_0=1/512$ 
for example  appears small in the absolute
scale, but
the gap to Fermi-energy ratio amounts 
to $\Delta_F/e_F \approx 0.4$ (see Table \ref{coherence-length}),
which is  larger than the
value $\Delta_F/e_F \approx 0.25$ at $\rho/\rho_0=1/8$ where the
gap is nearly the maximum. If we use
the Pippard's coherence length  $\xi_P=\hbar^2 k_F /m^{*}\pi\Delta_F$ in place
of the r.m.s. radius $\xi_{rms}$ (this may be justified at least for qualitative
discussion since the two quantities
agree within 10-25 \%, see Table \ref{coherence-length}), the ratio
between the r.m.s. radius and the average inter-neutron distance
is related to the gap to Fermi-energy ratio as
$\xi_{rms}/d \sim \xi_P/d \sim 0.2 e_F/\Delta_F$. 
Consequently we can expect in the zero-th order argument that 
the strong spatial correlation $\xi_{rms}/d \lesim 1$ 
emerges when the gap to Fermi-energy 
ratio is larger than $\Delta_F/e_F \gesim 0.2$.
This is realized in the present calculations in the density range
$\rho/\rho_0 \sim 10^{-4}-0.1$. In the following we shall investigate in detail
the spatial correlation in the neutron Cooper pair around this density range.

We shall comment here comparison with
the $^{3}SD_{1}$ neutron-proton pairing in symmetric nuclear
matter. In this case the BCS pairing gap calculated with a 
realistic bare force (the 
Paris force) is of the order of
8 MeV at maximum\cite{Garrido01}. The r.m.s. radius $\xi_{rms}$
of the neutron-proton Cooper pair is quite small, reaching to
the minimum value  $\xi_{rms} \sim 2$ fm at density around
$\rho/\rho_0 \sim 0.2$ \cite{Lombardo01a}.  Consequently, 
the r.m.s. radius $\xi_{rms}$ becomes considerably smaller
than the average inter-particle distance $d$ in the density interval
from $\rho/\rho_0 \sim 0.5$ down to the zero
density limit, where $\xi_{rms}$ becomes identical to the 
r.m.s. radius of the deuteron\cite{Lombardo01a}. 
In the case of the
$^{1}S$ neutron pairing, by contrast, the signature of the
strong coupling $\xi_{rms} \lesim d$ is obtained in the wide but
limited range of the density $\rho/\rho_0 \sim 10^{-4}-0.1$.
Apart from this difference, it is noted that the qualitative trend of
the Cooper pair size, e.g. shrinking
with increasing the density from the zero density limit, is similar to
that discussed in the neutron-proton case\cite{Lombardo01a}.

\begin{table}[htbp]
\begin{center}
\begin{tabular}{lccccccccccc}
\hline
\hline
      & $k_F$ & $\rho/\rho_0$ & $d$  & $m^*/m$  & $e_F$& $\Delta_F$ & 
$\xi_{rms}$  & $\xi_P$ & $P(d)$ &$(1/k_Fa)_\xi$& $(1/k_Fa)_\Delta$\\
\hline
\multicolumn{3}{l}{symmetric matter},  Gogny D1 &&&&&&&& \\
       & 1.36 & 1  & 2.27 & 0.668 &62.0& 0.64 & 46.60 & 41.76
 &0.18&-2.91&-2.99\\
       & 1.079 & 1/2  & 2.87 & 0.744 &33.8& 2.03 & 10.80 & 9.45
 &0.48&-1.83&-1.84\\
       & 0.68 & 1/8  & 4.55 & 0.891 &10.9& 2.60 & 4.81 & 3.87
 &0.81&-0.97&-0.90\\
       & 0.34 & 1/64  & 9.10 & 0.984 &2.45& 0.97 & 5.87 & 4.71
 &0.92&-0.59&-0.52\\
       & 0.17 & 1/512 & 18.20 & 0.998 &0.60& 0.22 & 12.05 & 10.30
 &0.91&-0.62&-0.58\\
\hline
\multicolumn{3}{l}{neutron matter}, G3RS   &&&&&&&& \\
       & 1.36 & 1    & 2.27 & 0.905 &42.4& 0.14&  159.8 & 144.1 &0.09
 &-3.70&-3.70\\
       & 1.079 & 1/2 & 2.87 & 0.925 &26.1& 1.52&  11.61 & 10.13 &0.47
 &-1.88&-1.85\\
       & 0.68 & 1/8  & 4.55 & 0.969 &9.89& 2.37&  4.94  & 3.90 &0.80
 &-0.99&-0.90\\
       & 0.34 & 1/64  & 9.10 & 0.995 &2.41& 0.98&  5.90 & 4.61 &0.92
 &-0.60&-0.50\\
       & 0.17 & 1/512 & 18.20 & 0.999 &0.60& 0.24& 11.16 & 9.30 &0.92
 &-0.55&-0.50\\
\hline
\hline
\end{tabular}
\end{center}
\caption{The pairing gap $\Delta_F$, the r.m.s. radius 
$\xi_{rms}$, the Pippard's coherence
length $\xi_P$, and the probability $P(d)$ within the average 
inter-neutron distance $d$,
associated with the neutron Cooper pair in symmetric nuclear matter
obtained with the Gogny D1 interaction and in
neutron matter with the G3RS force, at the neutron density 
$\rho/\rho_0=1, 1/2, 1/8, 1/64, 1/512$, or
equivalently $k_F=1.36, 1.094, 0.68, 0.34, 0.17$ fm$^{-1}$.
The Fermi energy $e_F$, the effective mass $m^*/m$, and the
average inter-neutron distance $d$ are also shown.
The parameters $(1/k_F a)_{\xi}$ and $(1/k_F a)_{\Delta}$ of the
regularized delta interaction model are also listed (see the text).
The units for $k_F$ and $\Delta_F, e_F$ are fm$^{-1}$ and MeV,
respectively while that for $d, \xi_{rms}$ and $\xi_{P}$ is fm.
\label{coherence-length}
}
\end{table}

\section{
Spatial structure of neutron Cooper pair}\label{spatial-sec}

\subsection{Cooper pair wave function: basics}\label{Cooper-wf-sec}

In examining the spatial structure of the neutron Cooper pairs,
we shall focus mostly
on the symmetric nuclear matter case obtained with the Gogny force and
the neutron matter case with the G3RS force.
The r.m.s. radii in these two cases represent 
a rough mean value of the four results plotted in Fig.\ref{rms}. Note also
that the two r.m.s. radii coincide with each other within $10\%$ in 
a very wide interval of density $\rho/\rho_0 = 10^{-3}- 0.5$.
It is by accident, but this feature can be exploited to 
single out influences of different interactions since the 
comparison can be made with the r.m.s. radii kept the same.

\begin{figure}[htbp]
\centerline{
\includegraphics[width=8.5cm]{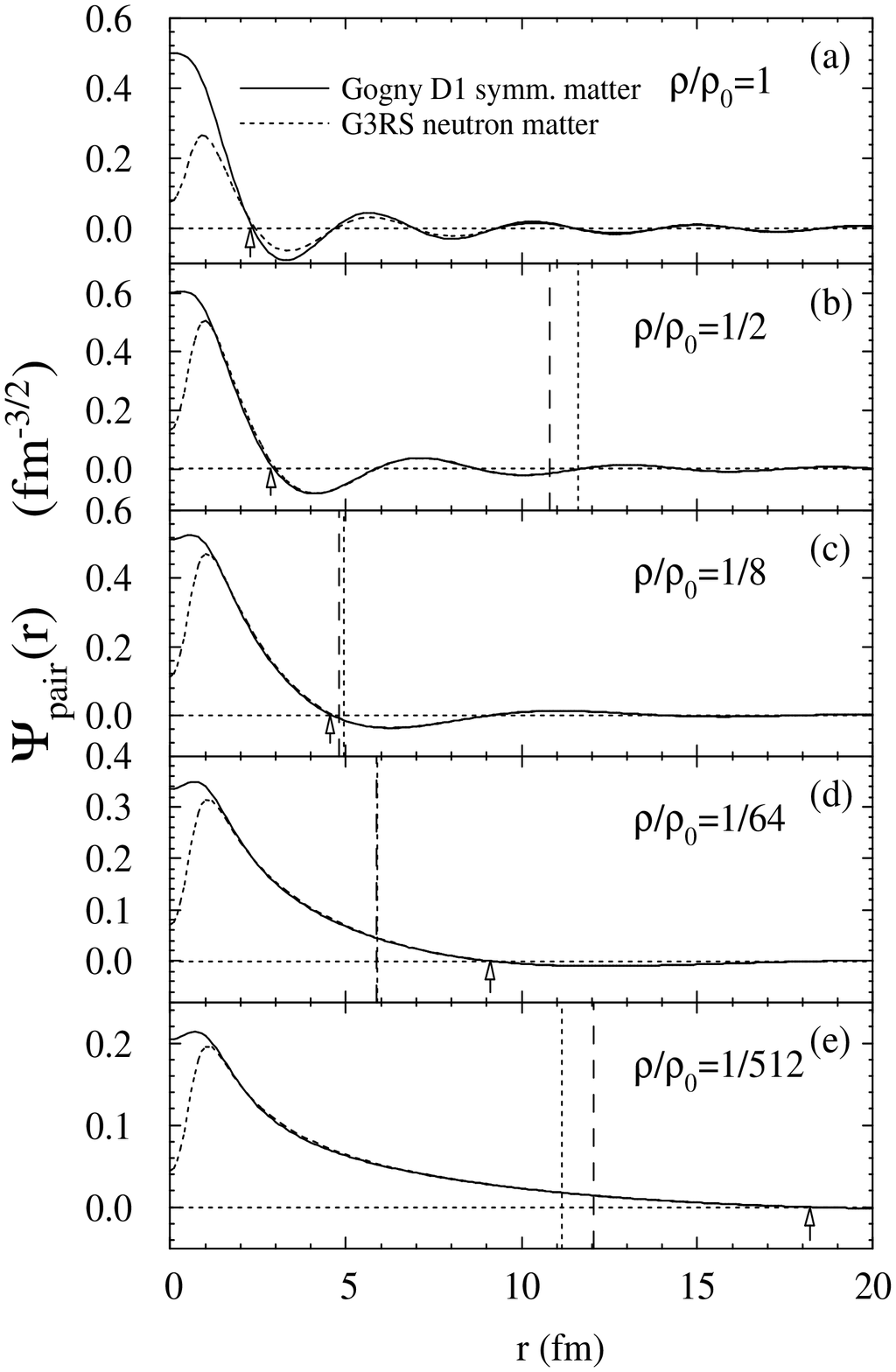}
\includegraphics[width=8.5cm]{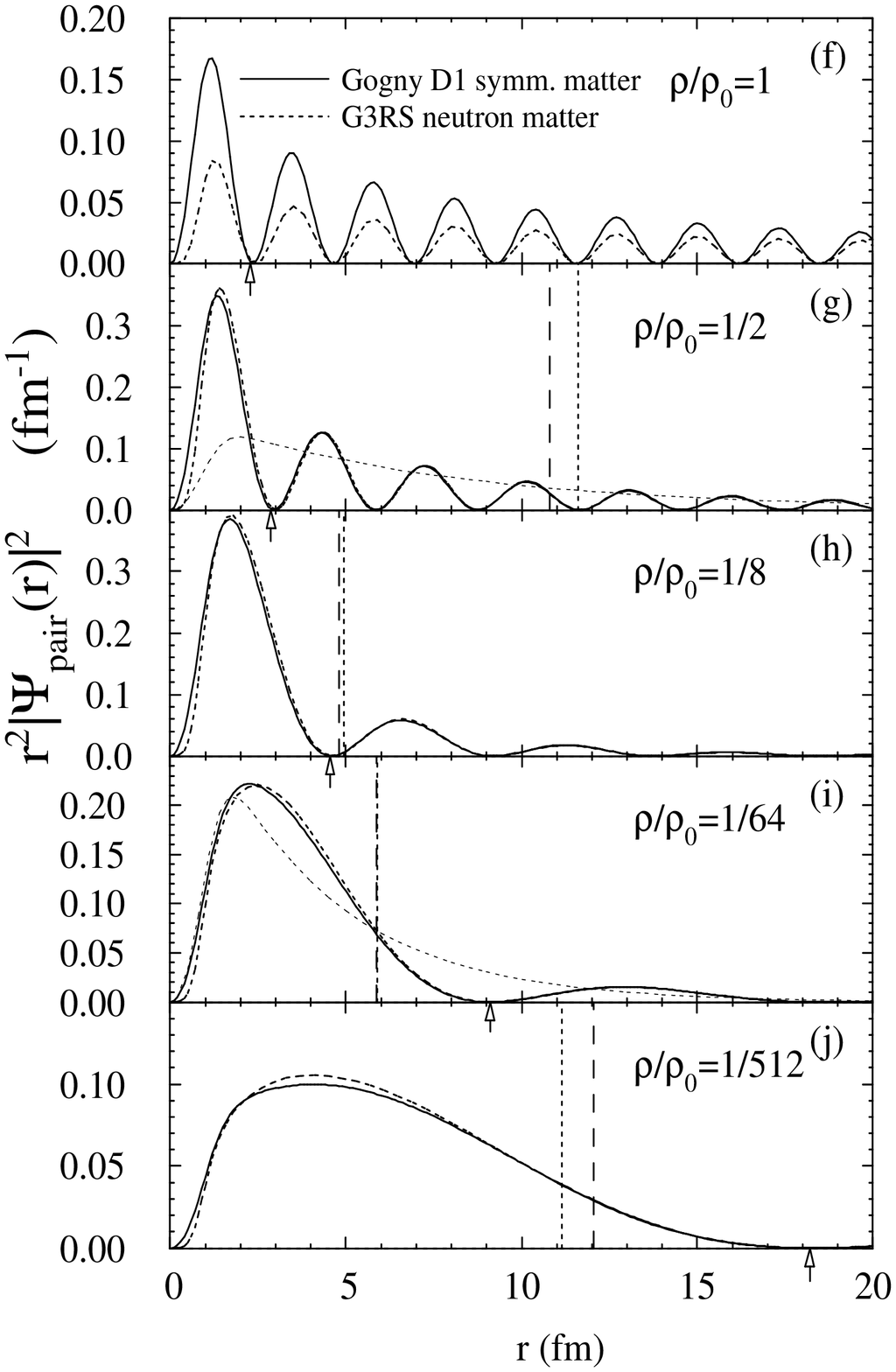}
}
\caption{(a-e)
The wave function $\Psi_{pair}(r)$ of the neutron Cooper pair as a function
of the relative distance $r$ between the pair partners at
the neutron density $\rho/\rho_0=1, 1/2, 1/8, 1/64, 1/512$.
The solid curve is for the pair in symmetric nuclear matter
obtained with the Gogny D1 force, while the dotted curve is for that
in neutron matter with the G3RS. 
The vertical dotted 
line represents the r.m.s. radius $\xi_{rms}$ of the Cooper pair in
neutron matter with the G3RS while the dashed line for
symmetric nuclear matter with Gogny D1. 
Here and also in the following figures the wave function 
is normalized by $\int_0^\infty |\Psi_{pair}(r)|^2 r^2 dr=1$.
The arrow indicates the average inter-neutron distance $d$.
(f-j) The same as (a-e) but for 
the probability density $r^2|\Psi_{pair}(r)|^2$.
The thin dotted line in (g) and (i) is the wave function of
the fictitious "bound state" in the free space described in the text. 
\label{Cooper-wf}}
\end{figure}

\begin{figure}[htbp]
\centerline{
\includegraphics[angle=0,width=8.5cm]{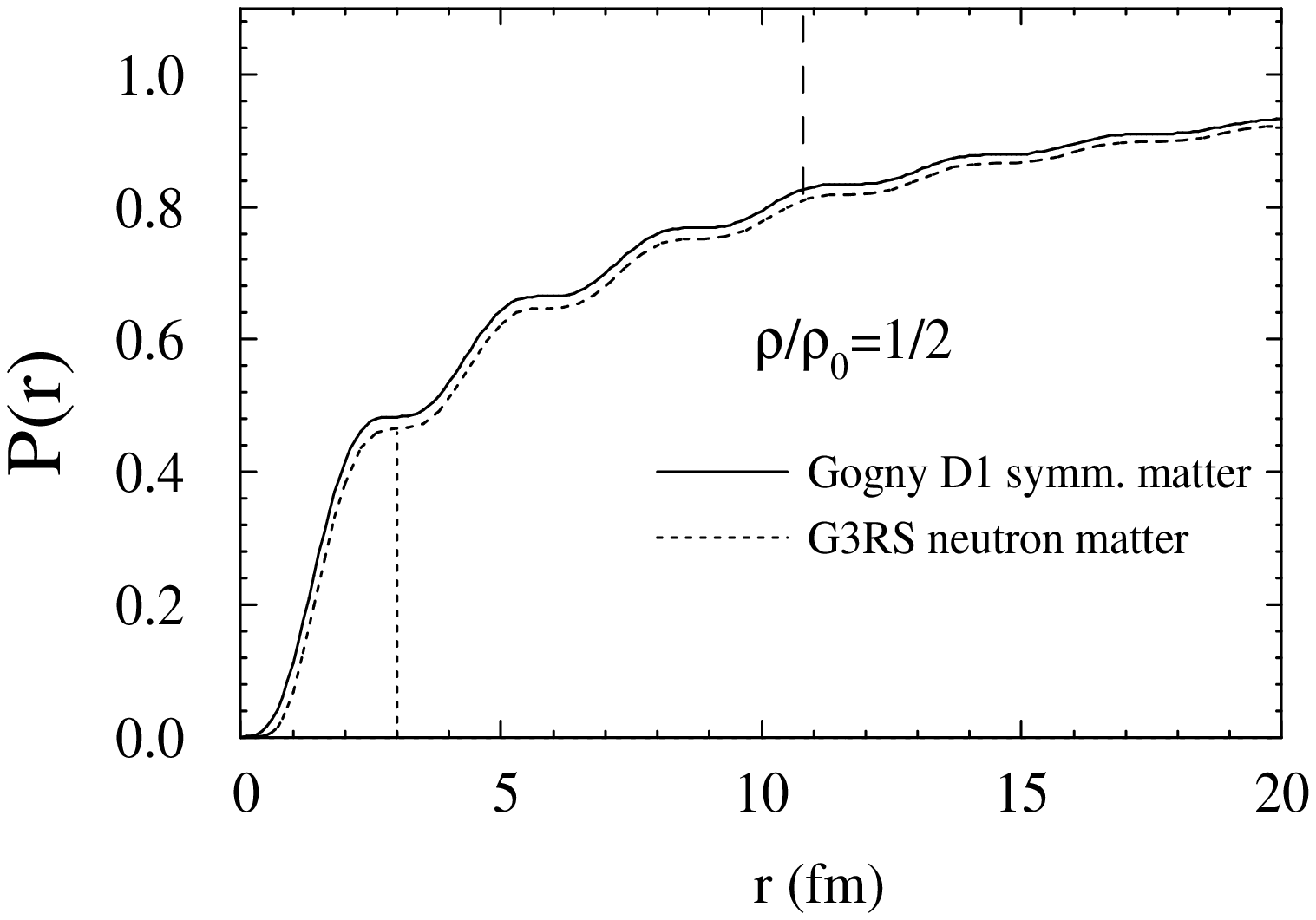}
}
\caption{
The probability $P(r)$ for the partner neutrons of the Cooper pair to 
be correlated 
within a relative distance $r$, calculated at density $\rho/\rho_0=1/2$.
The result for symmetric nuclear matter with the Gogny D1 force
is plotted with the solid curve while the dotted curve represents 
the result for neutron matter with the G3RS force.
The dashed vertical line indicates the r.m.s. radius of the
Cooper pair in the symmetric matter case. The vertical dotted line
marks the position $r=3$ fm. 
\label{pairprob}}
\end{figure}

Figures \ref{Cooper-wf}(a-e) show
the wave function $\Psi_{pair}(r)$ of the neutron Cooper pair
for the representative values of density listed in Table \ref{coherence-length}. 
The result for neutron matter 
calculated with the G3RS force and that for symmetric
nuclear matter with the Gogny D1 force are plotted in the same figure
for the reasons mentioned just above.
The
probability density $r^2|\Psi_{pair}(r)|^2$ multiplied 
by the volume element $r^2$ is plotted in Fig. \ref{Cooper-wf}(f-j).

As a quantitative measure of the spatial correlation,
we evaluate also the probability $P(r)$ for the partners of the
neutron Cooper pair to come close with each other within a
relative distance $r$. It is nothing but a partial
integration of the probability density $r^2|\Psi_{pair}(r)|^2$
up to the distance $r$:
\begin{equation}\label{prob-eq}
P(r) = {\int_0^{r} |\Psi_{pair}(r')|^2 r'^2 dr' \over
\int_0^{\infty} |\Psi_{pair}(r')|^2 r'^2 dr'}.
\end{equation}
An example of this quantity is shown in Fig.\ref{pairprob}
in the case of $\rho/\rho_0=1/2$.

Before proceeding to the main analysis we shall first 
point out that the G3RS and Gogny forces provide
essentially the same spatial structure of the Cooper pair except
at very short relative distances. 
In Fig.\ref{Cooper-wf}, a clear difference between the two forces 
is seen at short
relative distances $r \lesim 1$ fm.
(Note that the normal density case, Fig.\ref{Cooper-wf}(a,f), is
not relevant for this discussion since the gaps and the r.m.s. radii are
very different.) Apparently the suppression of the
wave function seen at $r \lesim 1$ fm in the G3RS case is 
caused by the strong repulsive core present in the bare force. 
The Cooper pair wave function for the Gogny force does 
not show this short range correlation because of the lack of the core.
On the other hand, by looking
at distances $r>1$ fm slightly larger than the core radius 
we find that
the Cooper pair wave functions obtained with the two 
forces agree quite well with each other. 
This observation applies also to the probability density
$r^2|\Psi_{pair}(r)|^2$ and the probability $P(r)$,
for which the difference at short distances $r<1$ fm becomes barely visible 
as the volume element is small at such short distances. Thus 
the spatial structure of the neutron Cooper pair does not depend on
whether the interaction is the bare force or the effective Gogny force,
provided that two cases gives the same r.m.s. radius of the Cooper pair.    
In the following, we shall
concentrate on behaviors which are common to the two interactions.

The Cooper pair wave function $\Psi_{pair}(r)$ in the coordinate representation
is reported in some of the previous BCS calculations adopting other models of 
the bare force\cite{Takatsuka84,Baldo90,Oslo96,DeBlasio97,Serra}
and the Gogny force\cite{Serra}.
Our wave function appears consistent with those in
Refs.\cite{Takatsuka84,Baldo90,Oslo96,Serra}
although 
in these references the wave function is shown
only at very limited numbers of density values and 
up to not very large relative distances.
 We found, however,
that the shape of the Cooper pair wave function shown in
Ref.\cite{DeBlasio97} differs largely from our results (Fig.\ref{Cooper-wf}),
especially at relative distances smaller than several fm.

\subsection{Density dependence}

If we compare in Fig.\ref{Cooper-wf} 
the Cooper pair wave functions at different density values,
important features show up.
An apparent observation is that
the spatial extension or the size of the Cooper pair
varies strongly with the density,  in accordance 
with the strong density dependence of the r.m.s radius $\xi_{rms}$ 
discussed in the previous section (Fig.\ref{rms}).   We emphasize here
another prominent feature. 
Namely {\it the profile} of the Cooper pair wave function
also changes significantly with the density.

At the normal density $\rho/\rho_0=1$ (Fig.\ref{Cooper-wf}(a,f)), the
Cooper pair wave function is spatially extended: 
the r.m.s. radius of the Cooper pair is 
as large as $\xi_{rms}\gesim 50$ fm.
The profile of the Cooper pair wave function in this case exhibits 
an exponential fall-off convoluted
with an oscillation. This behavior is 
consistent with the well known expression\cite{BCS}
$r\Psi_{pair}(r) \sim K_0(r/\pi \xi_P) \sin(k_F r)$ for the Cooper pair
wave function in the weak coupling BCS situation. Here $K_0$ is the
modified Bessel function, which behaves asymptotically as 
$K_0(r/\pi \xi_P) \sim (\xi_P/r)^{1/2}\exp(-(r/\pi\xi_P))$.
The position of the first node 
$r \approx \pi k_F^{-1}$ approximately 
corresponds to the average inter-particle
distance $d=3.09k_F^{-1}(=2.3 {\rm fm})$.
The wave function has significant amplitude for $r>d$ since we here have
a relation $\xi_{rms,P} \gg d$ (see Table \ref{coherence-length}). 
This is a typical behavior in the situation of the
weak coupling BCS.

\begin{figure}[htbp]
\centerline{
\includegraphics[angle=270,width=8.5cm]{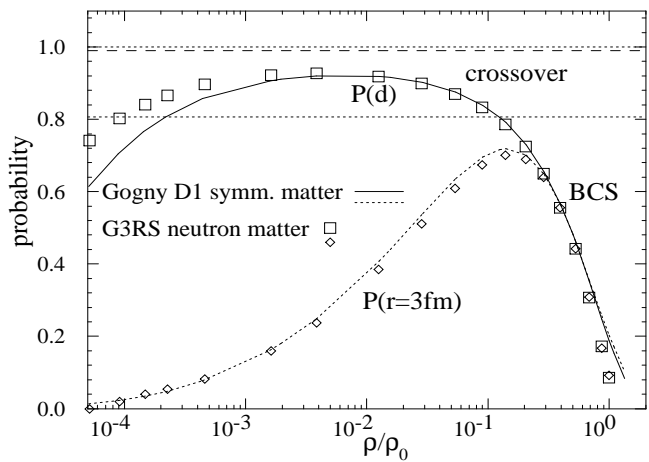}
}
\caption{
The probability $P(d)$ for the partner neutrons to be correlated within
the average inter-neutron distance $d$ and the
probability $P(3{\rm fm})$ within $r=3$ fm. The solid and dotted curves are
for symmetric nuclear matter obtained with the Gogny D1
force, while the square and diamond symbols are for neutron matter with
 the G3RS
 force.  
\label{prob}}
\end{figure}

The Cooper pair wave function at the density $\rho/\rho_0=1/8$
(Fig.\ref{Cooper-wf}(c,h)) is very
different from that at the normal density. 
Apart from the considerably small spatial extension 
($\xi_{rms}= 4.8-4.9$ fm), the
functional form of the wave function behaves quite differently. 
We find that
amplitude of the wave function is strongly concentrated within
the average inter-neutron
distance $d$, and that the oscillating amplitude beyond $d$ is quite small.
This is consistent with the observation in the previous section that 
the r.m.s. radius $\xi_{rms}$ of the Cooper pair 
is smaller than the average inter-neutron distance $d$ in this case
(cf.Fig.\ref{rms} and Table \ref{coherence-length}).  
The probability $P(d)$  
for the partners of the Cooper pair to be correlated 
within the inter-nucleon distance
$d$, 
exceeds $0.8$ (Fig.\ref{prob} and Table \ref{coherence-length}), 
indicating directly the strong spatial di-neutron correlation.

The neutron Cooper pair wave functions 
at $\rho/\rho_0=1/64$ and  1/512,  Fig.\ref{Cooper-wf}(d,e,i,j), exhibit a
behavior similar to that at $\rho/\rho_0=1/8$. 
Inspecting more closely, we notice that
the concentration within $r<d$ is stronger than at $\rho/\rho_0=1/8$ while
the spatial extension
itself is slightly larger ($\xi_{rms} \approx 6-12$ fm). 
We observe also
smaller oscillating amplitude 
in the large distance region $r > d$ (Fig.\ref{Cooper-wf}), 
larger values of $P(d)\approx 0.9$, and smaller ratio $\xi_{rms}/d$
(Table \ref{coherence-length}).  
They all point to 
stronger spatial di-neutron
correlation at these values of density.

The Cooper pair wave function
at $\rho/\rho_0 = 1/2$ (Fig.\ref{Cooper-wf}(b,g)) exhibits 
an intermediate feature 
between that at the normal density $\rho/\rho_0=1$ and those at
$\rho/\rho_0 =1/64 - 1/512$.  In particular, we notice  that
the spatial correlation seen at the lower density persists 
to some significant extent also in this case.
For example,
the probability density 
is strongly concentrated to the short distance region of
$r \lesim 3$ fm (Fig.\ref{Cooper-wf}(g)). This is more apparent in
the plot of $P(r)$ shown in 
Fig.\ref{pairprob}, where we find that
the probability $P(r)$ increases steeply with increasing $r$
from $r=0$, and
reaches $\sim 50\%$ already at $r=3$ fm, which 
is roughly the interaction range of the nucleon force.
This strong concentration within $r<3$ fm 
may be elucidated by comparing with
what could be
expected if a bound pair having 
the same r.m.s. radius ($\xi_{rms}=10.8$ fm in this case) was formed 
in the free space. 
(We calculate this fictitious ``bound state'' wave function by increasing the
strength of the Gogny D1 potential by a numerical factor.)
It is noticed that the profile of the Cooper pair wave function
differs from the ``bound state'' wave function which is plotted 
with the thin dotted line 
in Fig.\ref{Cooper-wf}(g).
In this "bound state" wave function, 
concentration of the probability within $r \lesim 3$ fm 
is not very large, i.e., $P(3{\rm fm})=0.24$,
while the probability $P(3{\rm fm})$ associated with the neutron Cooper 
pair wave function is about twice this value. 
This indicates 
that the spatial
di-neutron correlation is also strong for the moderate low density region  
$\rho/\rho_0\sim 0.5$.  
A remnant of this spatial correlation is found also at the normal
density (Fig.\ref{Cooper-wf}(a,f)), 
but in this case the concentration within the interaction range
is not very large ($P(3{\rm fm})=0.21$), 
due to the very large Cooper pair size 
( $\xi_{rms} \sim 50$ fm). 
In contrast the Cooper pair wave function at the lower density, 
$\rho/\rho_0=1/64$ (Fig.\ref{Cooper-wf}(i)) 
for example, 
is much more similar to the "bound state" wave function.

Figure \ref{prob} shows the  overall behavior of 
$P(3{\rm fm})$  and $P(d)$ as a function of the density. It is seen
that the strong concentration
within the
interaction range, say $P(3{\rm fm})>0.5$, 
is realized in
the density region $\rho/\rho_0 \approx 5\times 10^{-2}-0.5$.
The probability $P(3{\rm fm})$ reaches the maximum value 
$\sim 0.7$ around $\rho/\rho_0 \sim 0.1$, where the r.m.s.
radius is the smallest. At lower density
$\rho/\rho_0 \lesim 10^{-1}$, the probability 
$P(3{\rm fm})$ decreases gradually in accordance with the gradual increase
of the r.m.s. radius of the Cooper pair. Note however that in this density region
($\rho/\rho_0 \sim 10^{-4}-10^{-1}$) the concentration of the probability 
within
the average inter-neutron distance $d$ remains very large, i.e., $P(d) \gesim 0.8$.

All the above analyses indicate that the
spatial di-neutron correlation is strong in the quite wide density interval
$\rho/\rho_0 \sim 10^{-4} - 0.5$.

It is interesting to compare our result with that in a similar analysis 
of the Cooper pair wave
function for the $^3SD_1$ neutron-proton pairing. In that case
the Cooper pair wave function is found to merge smoothly into the
deuteron wave function in the low density limit\cite{Baldo95}.
Correspondingly, 
the r.m.s. radius of the Cooper pair approaches to that of the deuteron,
which is much smaller than the average 
inter-particle distance\cite{Lombardo01a}.
This is interpreted as a realization of 
the BEC of the deuterons in the low density
region and the BCS-BEC crossover taking place with change of the
density\cite{Stein95,Baldo95,Lombardo01a,Lombardo01b}. In the
neutron pairing case, the similarity of the Cooper pair
wave function to a bound state wave function is found only  
in a limited density range 
$\rho/\rho_0 \sim 10^{-4}-10^{-1}$, and it never converges 
to a bound state wave function (NB. there is no bound state in this
channel in the free space).
The r.m.s. radius $\xi_{rms}$ of the Cooper pair is only comparable
to the average inter-neutron distance $d$ in the same density interval.
Although the spatial correlation is strong as discussed above, 
these
qualitative observations alone are not enough to assess 
whether the region of the BCS-BEC crossover 
is reached in the case of the neutron pairing.
We shall investigate this issue on more quantitative bases in the next section.

\section{Relation to the BCS-BEC crossover}\label{BCS-BEC-sec}

In the previous section, we have seen the strong spatial
correlation in the neutron Cooper pair wave function at low density. 
In the present section we shall elucidate its
implication by making a connection
to the BCS-BEC crossover phenomenon.

For this purpose, we shall first describe a reference Cooper pair
wave function based on a simple
solvable model of the BCS-BEC crossover, and then compare our results with that of the reference model.
As such a reference, we adopt a $^{1}S$ pairing model that applies generically to
a dilute gas limit of any Fermion systems\cite{Leggett,Melo,Engelbrecht}, 
for which the average inter-particle distance 
$d=\rho^{1/3}$ is supposed to be much larger 
than the range of interaction.
The dilute limit is equivalent to treating 
the interaction matrix elements
as a constant, or assuming a contact interaction. 
Using the relation between the interaction constant 
and the zero-energy $T$-matrix or the scattering length $a$,
the gap equation (\ref{gap-eq}) is written in a regularized form:
\begin{equation}\label{reg-gap-eq}
{m \over 4\pi \hbar^2 a }= - {1 \over 2(2\pi)^3}\int d \veck 
\left({1 \over E(k)} - {1 \over e(k)}\right).
\end{equation} 
The regularized gap equation (\ref{reg-gap-eq}) and the number equation (\ref{num-eq}) 
are now expressed 
analytically in terms of some special functions,
and are easily solvable\cite{Marini,Papenbrock}.
In this model a dimensionless parameter
$1/k_F a$, characterized by the scattering length and the Fermi momentum, 
is the only parameter that controls the
strength of interaction, and hence properties of the pair correlation
are determined solely by $1/k_F a$ while the length scale is given by
$k_F^{-1}$  (or the inter-particle distance $d=3.09k_F^{-1}$), and 
the energy scale  by the Fermi energy $e_F= \hbar^2 k_F^2 /2m$
\cite{Leggett,Melo,Randeria,Engelbrecht,Marini}. 
The gap to Fermi-energy ratio $\Delta/e_F$ and the ratio $\xi_{rms}/d$ 
between the r.m.s. radius and the average inter-particle distance
are then monotonic functions of the 
interaction parameter $1/k_F a$\cite{Engelbrecht,Marini}. The functional form
of the Cooper pair wave function, defined by Eq.(\ref{Cooper-eq}), 
is also 
determined only by $1/k_F a$, except for the length scale.
Since there is no available analytic expression for the wave function,
we evaluate Eq.(\ref{Cooper-eq}) by performing the momentum integral 
numerically with a help of an explicit use of a smooth cut-off 
function of a Gaussian form. 
The cut-off scale is chosen large enough so that the results shown below
do not depend on it.

The range $1/k_F a \ll -1$ of the interaction parameter corresponds to 
the situation of the weak coupling BCS, for which 
the pairing gap is given by
the well known formula\cite{Leggett,Randeria,Lombardo-Schulze}
$\Delta/e_F \approx 8e^{-2}\exp\left(\pi / 2k_F a\right)$.
In the opposite range $1/k_F a \gg 1$, the 
situation of the Bose-Einstein condensation (BEC) of bound
Fermion pairs (bosons) is realized. 
The crossover between the weak coupling BCS and the strong coupling
BEC corresponds to the interval 
$-1 \lesim 1/k_F a \lesim 1$, as described in Refs.
\cite{Leggett,Melo,Randeria,Engelbrecht}. 
(The case with the infinite scattering
length $1/k_F a=0$ is the midway of the
crossover, called the unitarity limit.) 
In the following we shall adopt $1/k_F a=\pm 1$ 
according to Ref.\cite{Engelbrecht}
as boundaries characterizing the crossover 
although the transition is smooth in nature.

\begin{figure}[htbp]
\centerline{
\includegraphics[angle=270,width=5cm]{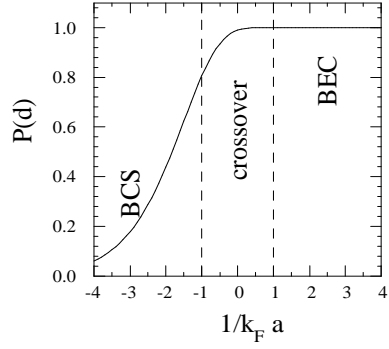}
}
\caption{
The probability $P(d)$ for the partner particles to be correlated
within the average inter-particle distance $d$
for the regularized delta interaction model.
\label{prob-delta}}
\end{figure}

In Table \ref{BCS-BEC} we list 
the values of $\xi_{rms}/d$
and $\Delta_F/e_F$ at the ``boundaries'' 
$1/k_Fa= \pm 1$ of the crossover domain and 
at the unitarity limit $1/k_Fa=0$ \cite{Engelbrecht,Marini}.
Note that the r.m.s. radius comparable to the average inter-particle
distance $0.2  \lesim \xi_{rms}/d \lesim 1.1$, or the
gap comparable to the Fermi energy 
 $0.2 \lesim \Delta_F/e_F \lesim 1.3$ corresponds to 
the BCS-BEC crossover domain $-1 \lesim  1/k_F a \lesim 1$.
We have discussed in the previous section 
the probability $P(d)$ for the paired neutrons
to come closer than the average inter-neutron distance $d$ 
(cf. Fig.\ref{prob}). As the same quantity is easily calculated
also in the analytic model of the BCS-BEC crossover 
(the result is shown in Fig.\ref{prob-delta}), this quantity
may be used also as a measure of the crossover. 
The calculated boundary values corresponding to $1/k_F a = \pm1, 0$ 
are listed in Table \ref{BCS-BEC}.
The crossover region is specified 
by $0.8 \lesim P(d)\lesim 1.0$ while the boundary to the strong 
coupling BEC regime ($P(d) \approx 1$) is hardly visible in this measure.

In the case of nucleonic matter, the assumption of 
the dilute gas limit 
may be justified only at very low density $\rho/\rho_0 \lesim 10^{-5}$ 
(or $k_F \lesim 0.05$ fm$^{-1}$)\cite{Heiselberg01,Khodel,Lombardo-Schulze},
and hence we cannot apply 
the above analytic model in a direct manner to the region of the
density $\rho/\rho_0=10^{-5}-1$ which we are dealing with. 
In order to make the application possible, we shall stand
on a more flexible viewpoint by
regarding the interaction parameter $1/k_F a$ as a freely adjustable
variable, rather than by fixing it from the physical value of the 
neutron scattering length. We shall call the model treated in this way
the regularized delta interaction model to distinguish from
the original idea of the dilute gas limit.

\begin{figure}[htbp]
\centerline{
\includegraphics[width=8.5cm]{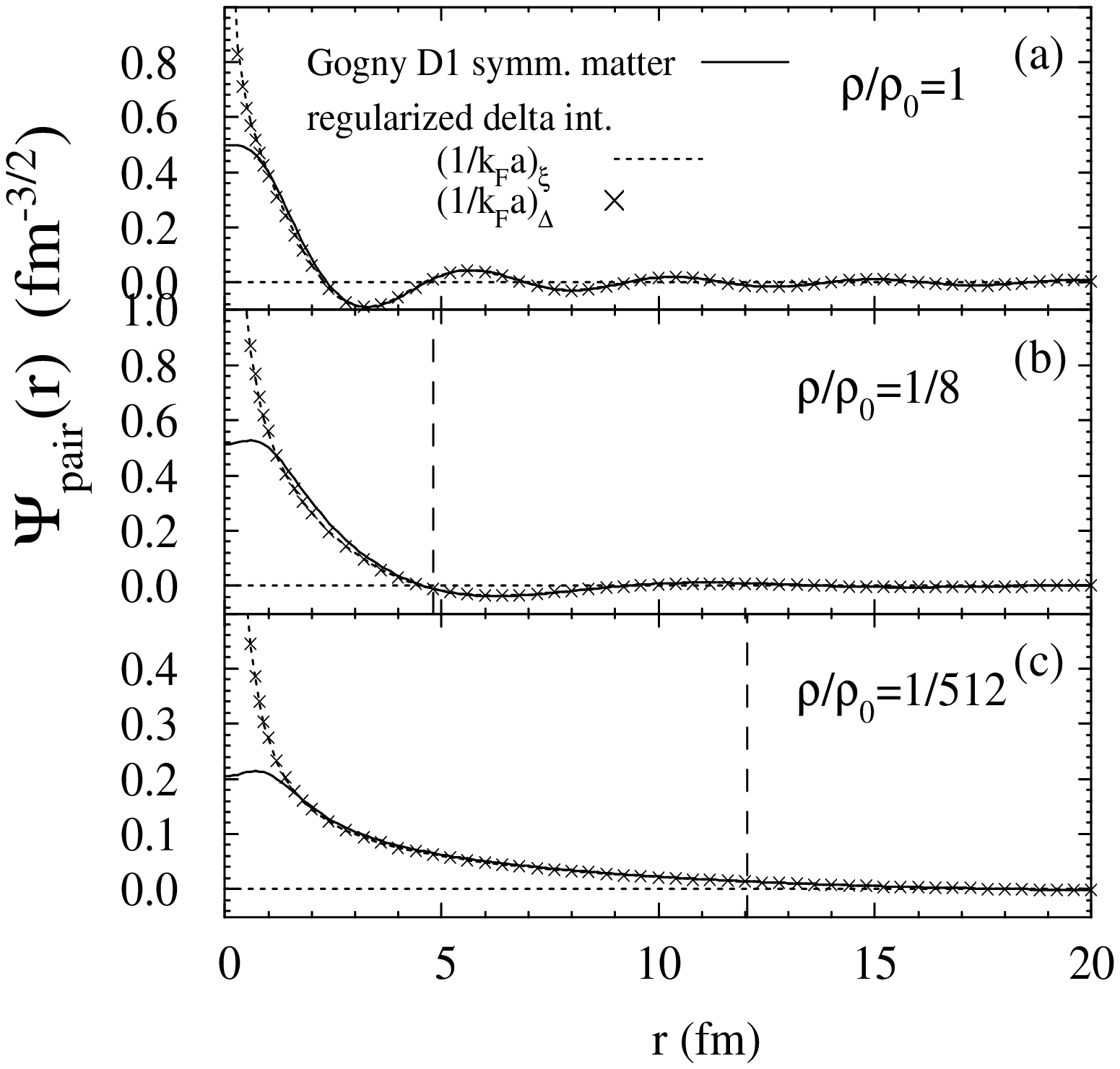}
\includegraphics[width=8.5cm]{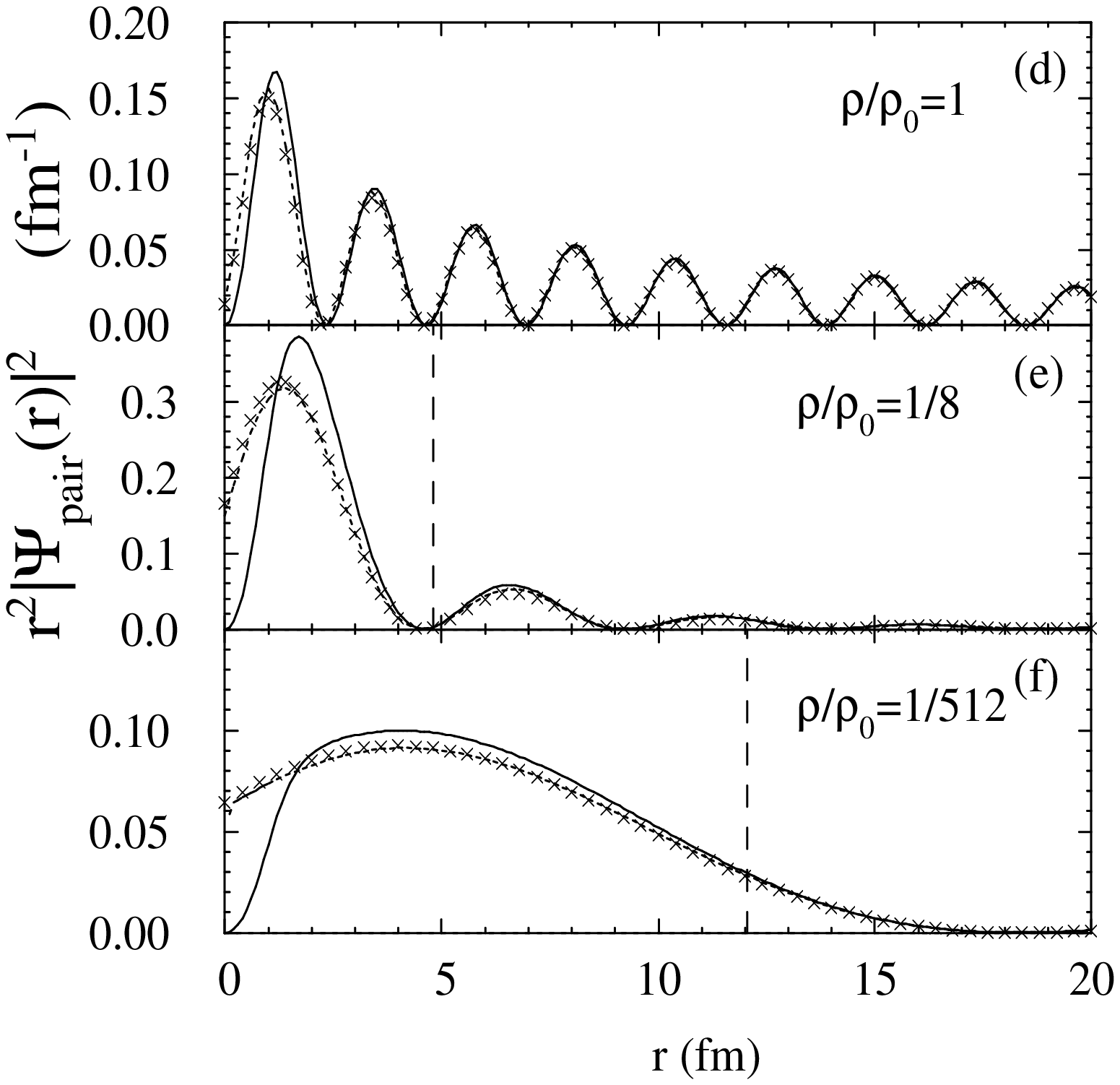}
}
\caption{(a-c)
The neutron Cooper pair wave function $\Psi_{pair}(r)$
in the regularized delta interaction model, plotted
with the dotted curve and the cross symbol
in the cases of 
$(1/k_F a)_{\xi}$ and  $(1/k_F a)_{\Delta}$, respectively. 
The neutron Cooper pair wave function in
symmetric nuclear matter obtained with the Gogny D1 force
is also shown by the solid curve.
(d-f) 
The same as (a-c) but for the probability density $r^2|\Psi_{pair}(r)|^2$.
\label{Cooper-wf-delta}}
\end{figure}

The interaction parameter $1/k_F a$ needs to be determined then.
We shall require the condition that the regularized delta interaction
model gives, for a given value of density, the same r.m.s.
radius $\xi_{rms}/d$ 
as that of the microscopically calculated neutron Cooper pair.
The parameter determined in this way may be denoted $(1/k_Fa)_{\xi}$. 
We can also determine  the interaction parameter to 
reproduce the ratio $\Delta_F/e_F$ between the gap and the
Fermi energy, which we shall denote $(1/k_Fa)_{\Delta}$.  The values of 
$(1/k_Fa)_{\xi}$ and $(1/k_Fa)_{\Delta}$ thus determined are listed in 
Table \ref{coherence-length}. There is no sizable difference
between $(1/k_Fa)_{\xi}$ and $(1/k_Fa)_{\Delta}$.
The Cooper pair wave 
functions obtained in this reference model are shown in
Fig.\ref{Cooper-wf-delta}.
It is hard to distinguish between the two options of
$1/k_F a$. We now compare them
with the neutron Cooper pair obtained with the
Gogny force for symmetric matter at the three representative
values of density $\rho/\rho_0=1, 1/8$ and 1/512.

It is seen from Fig.\ref{Cooper-wf-delta} that 
the wave function of the regularized delta interaction model and 
and that of the neutron 
Cooper pair behave very similarly at
distances far outside the interaction range, $r \gesim 5$ fm. 
In contrast, we notice a 
sizable disagreement for $r \lesim 3$ fm.  
The disagreement is understandable
as the wave function within the interaction range $r \approx 3$ fm 
of the finite range Gogny force could not be 
described by the zero-range delta interaction.
The Cooper pair wave function in the regularized delta interaction model
exhibits the known divergence $\Psi_{pair}(r) \propto 1/r$ for 
$r \rightarrow 0$ (cf. Fig.\ref{Cooper-wf-delta}(a-c)), and consequently 
the disagreement between the Gogny model and the regularized 
delta interaction model
becomes serious at very short relative distances
$r \lesim 1$ fm.
Note however that the squared wave function weighted with the 
volume element $r^2$ stays finite as seen Fig.\ref{Cooper-wf-delta}(d-f)
and hence there is no diverging difference in the probability
density. 
These observations suggest that 
the regularized delta interaction model 
can account for the essential features of the spatial structure of the 
neutron Cooper pair as far as the interaction strength
$1/k_F a$ is chosen appropriately. 

\begin{table}[htbp]
\begin{center}
\begin{tabular}{ccccc}
\hline
\hline
$1/k_F a$  &  $\xi_{rms}/d$  & $\Delta/e_F$ & $P(d)$  & \\
\hline
 -1  &  1.10   &   0.21   &  0.807 &  boundary to BCS \\
  0  &  0.36   &   0.69   &  0.990 &  unitarity limit \\
  1  &  0.19   &   1.33   &  1.000 &  boundary to BEC \\
\hline
\hline
\end{tabular}
\end{center}
\caption{The reference values of 
$1/k_F a$, $\xi_{rms}/d$  and $\Delta/e_F$ 
characterizing the
BCS-BEC crossover in the regularized delta interaction model.}
\label{BCS-BEC}
\end{table}

\begin{figure}[htbp]
\centerline{
\includegraphics[angle=270,width=8.5cm]{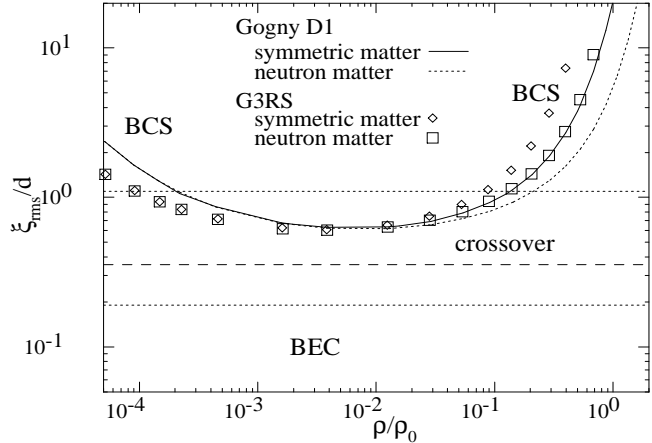}
}
\caption{
The ratio $\xi_{rms}/d$ between the r.m.s. radius $\xi_{rms}$ of the neutron
Cooper pair and the average inter-nucleon distance $d$, calculated 
with the Gogny D1 force for
symmetric nuclear and neutron matter 
(the solid and dotted lines, respectively),
and those  with the G3RS force for symmetric nuclear
and neutron matter (the diamond and square
symbols), plotted as a function of the neutron density.
The reference values
characterizing the BCS-BEC crossover listed in Table \ref{BCS-BEC}
are also shown with  the horizontal 
dotted and dashed lines.
\label{rmsoverd}}
\end{figure}

We thus have a reference frame, i.e., 
the regularized delta interaction model,
to which the neutron pairing is mapped. The question on possible relation
to the BCS-BEC crossover phenomena can be addressed now quantitatively.
We first look into
the ratio $\xi_{rms}/d$ between the r.m.s radius $\xi_{rms}$ of 
the Cooper pair and the average inter-particle distance $d$.
The values of $\xi_{rms}/d$ for the neutron
Cooper pair obtained with the G3RS force and the Gogny D1 interaction
are compared in Fig.\ref{rmsoverd} with the reference values 
defining the ``boundaries'' of the BCS-BEC crossover domain
(Table \ref{BCS-BEC}).
It is seen in
Fig.\ref{rmsoverd} that
the calculated ratio $\xi_{rms}/d$ enters the domain 
of the BCS-BEC crossover,  $1.10 > \xi_{rms}/d (>0.19) $,
in the density interval $\rho/\rho_0 \approx  10^{-4} - 0.1$. 
Note also the calculated ratio becomes  closest to
the unitarity limit $\xi_{rms}/d=0.36$ around the density 
$\rho/\rho_0 \sim 10^{-2}$. 
In the other way around, the weak coupling BCS regime is realized
only at very low density $\rho/\rho_0 \lesim 10^{-4}$ and 
around the normal density $\rho/\rho_0 \gesim 0.2$.

\begin{figure}[htbp]
\centerline{
\includegraphics[angle=270,width=8.5cm]{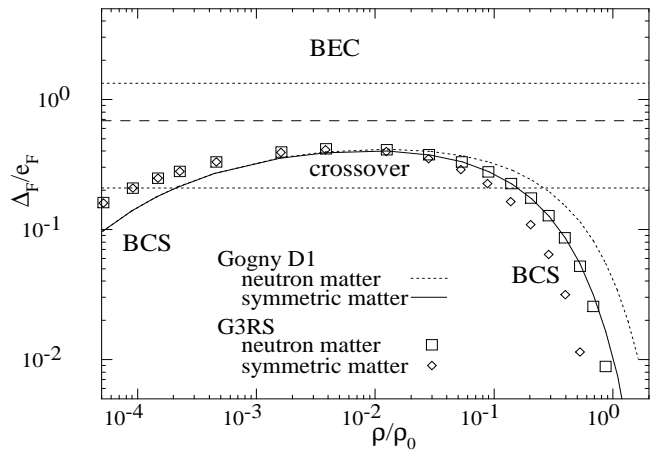}
}
\caption{
The ratio $\Delta_F/e_F$ between the neutron pairing gap $\Delta_F$
and the neutron Fermi energy $e_F$, plotted as a function of
the neutron density. The curves and the symbol are the same as in
 Fig.\ref{rmsoverd}.
The reference values  
characterizing the BCS-BEC crossover listed in Table \ref{BCS-BEC}
are also shown with the horizontal 
lines.
\label{deltaoveref}}
\end{figure}

Comparing in Fig.\ref{deltaoveref}
the gap to Fermi-energy ratio $\Delta_F/e_F$ with the 
boundary values $0.21<\Delta_F/e_F<1.33$, 
we have the same observation that
the density region $\rho/\rho_0 \sim 10^{-4}-0.1$ corresponds to
the domain of the BCS-BEC crossover. Comparison of
the third measure $P(d)$, performed in Fig.\ref{prob}, provides
us the same information. 
It is seen also 
in the values of $(1/k_Fa)_\xi$ and $(1/k_Fa)_\Delta$
listed in Table \ref{coherence-length} that
the condition of the crossover region $(1/k_Fa)_{\xi,\Delta}>-1$
is met in the cases of $\rho/\rho=1/8,1/64$ and 1/512.

On the basis of the above analysis, we conclude that the
strong spatial correlation at short relative distances 
seen in the neutron Cooper pair 
in the very wide density range 
$\rho/\rho_0 \approx 10^{-4} - 0.1$ is the behavior associated
with the BCS-BEC crossover.

\section{density-dependent delta interaction}\label{DDDI-sec}

\subsection{DDDI and the cut-off energy}\label{cutoff-sec}

The contact force whose interaction strength is chosen
as a density-dependent parameter, called often the density-dependent delta
interaction (DDDI), has been employed as a phenomenological
effective interaction describing the pairing correlation in finite nuclei,
especially unstable nuclei with large neutron excess
\cite{DobHFB2,DD-Dob,DD-mix,Matsuo05,YuBulgac,Grasso,Stoitsov03,Teran,Bender}.
In the previous section we found that 
the regularized contact interaction model, which employs also the 
contact force but with an analytic regularization,
describes the essential feature of the spatial structure of the 
neutron Cooper pair wave function in the whole density range. 
This suggests a possibility that 
the phenomenological DDDI may also describe the
spatial correlation in the neutron pairing.
We would like to examine from this viewpoint in what conditions 
the DDDI can be justified. 

The density-dependent delta interaction has a form
\begin{equation}\label{dddi-eq}
v(\vecr)={1-P_\sigma \over 2}V_0[\rho]\delta(\vecr),
\end{equation}
where $V_0[\rho]$ is the interaction strength which is supposed to be
dependent on the density. The force acts only in the
$^{1}S$ channel due to the projection operator $(1-P_\sigma)/2$.
It should be noted that in the applications of the density-dependent
delta interaction to finite nuclei, 
an explicit and finite cut-off energy 
needs to be introduced. 
The cut-off energy in this case is regarded as an additional model parameter.

Applying the DDDI 
to the neutron pairing in uniform matter, 
the gap equation reads 
\begin{equation}\label{gap-dddi-eq}
\Delta= - {V_0 \over 2(2\pi)^3}\int' d \veck 
{\Delta \over E(k)}
\end{equation}
where the pairing gap here is a momentum independent constant $\Delta$.
For the single-particle energy $e(k)$,
we adopt the effective mass approximation. 
We use the effective mass derived from the Hartree-Fock spectrum
of the Gogny D1.
The momentum integral in Eq.(\ref{gap-dddi-eq}) is performed
under a sharp cut-off condition
\begin{equation}\label{ecut}
e(k) < \mu + e_{cut}
\end{equation}
where we define the cut-off energy $e_{cut}$ as relative energy
from the chemical potential $\mu$.  
We shall treat $e_{cut}$ as
a common constant which is applied to all density.
We use the same cut-off in calculating the Cooper pair wave function.

We remark that our definition of the cut-off 
is different from a similar one adopted in 
Refs.\cite{Bertsch91,Garrido},
where a cut-off
energy is defined  with respect to the
single-particle energy $e(k)$ measured from the bottom of
the spectrum $e(k=0)$, e.g., by imposing $e(k) < 60$ MeV\cite{Garrido}
independent of the density. In our case, by contrast, we fix
an energy window above the chemical potential $\mu$ to $e_{cut}$.
We think that the cut-off energy $e_{cut}$ defined in this way can be 
compared with the quasiparticle energy cut-off $E_i <E_{cut}$ adopted often
in the HFB calculations for finite nuclei ($E_i$ is the quasiparticle energy
of the single-particle state $i$)
\cite{DobHFB,DobHFB2,DD-Dob,DD-mix,Matsuo05,YuBulgac,Grasso,Stoitsov03,Teran,Bender}. 
Note that in finite nuclei the density varies
locally with the position coordinate while the
quasiparticle energy is defined globally. This may imply, in the
sense of the local density approximation, that a fixed density-independent
cut-off quasiparticle energy $E_{cut}$ is applied to each value of 
local density. 
Note also that both our $e_{cut}$ and the cut-off quasiparticle energy
$E_{cut}$ are quantities measured from the chemical potential, 
which approximately coincide
if $e_{cut}$ and $E_{cut}$ are sufficiently larger than the pairing gap.
Thus we can compare directly $e_{cut}$ and $E_{cut}$, provided that 
$e_{cut}$ is chosen as a density-independent constant.
At the zero-density limit with $\mu=0$,
the cut-off energy $e_{cut}$ simply defines an upper bound on the
free single-particle energy $e(k)= \hbar^2 k^2/2m$.

\subsection{Constraints on $e_{cut}$}\label{cutoffdep-sec}

Let us 
investigate how the energy cut-off influences the Cooper pair wave function.
We performed several calculations using different values of 
$e_{cut}=5,10,30, 50, 100, 200$ MeV for symmetric nuclear matter. 
In doing so, we choose
the interaction strength $V_0$ for each value of density so that the 
gap $\Delta$ calculated with a given cut-off energy
$e_{cut}$ coincides with the gap $\Delta_F$ obtained 
with the Gogny D1 force. Note that the
interaction strength $V_0$ thus determined 
depends on both the cut-off energy 
and the density.

\begin{table}[htbp]
\begin{center}
\begin{tabular}{cccccccc}
\hline
\hline
    &\multicolumn{6}{c}{$\xi_{rms}$ [fm]} &  \\
$\rho/\rho_0$  &  $e_{cut}=$5 &  10  & 30  & 50  & 100 & 200 MeV & Gogny D1\\    
\hline
1     & \underline{48.6}  & \underline{47.3} & \underline{46.7} & \underline{46.6} & \underline{46.5} & \underline{46.5}
 & 46.6 \\
1/2    & 16.9  &  13.8 & \underline{11.2} & \underline{10.9} & \underline{10.7} & \underline{10.6} &
 10.8  \\
1/8     & 14.9  &  10.0 &  6.0 & \underline{5.3} & \underline{4.8} & \underline{4.6} &  4.8 \\
1/64   &  10.5 &  7.9  & \underline{6.2}  & \underline{6.0} & \underline{5.7} & \underline{5.5} &  5.9
 \\
1/512  & \underline{13.2}  &  \underline{12.5} & \underline{12.0} & \underline{11.9} & \underline{11.8} & \underline{11.7} &  12.1 \\
\hline
\hline
\end{tabular}
\end{center}
\caption{The r.m.s. radius $\xi_{rms}$  of the neutron 
Cooper pair in symmetric nuclear matter obtained with the
density dependent delta interaction 
and different cut-off energies 
$e_{cut}=5,10,30,50,100,200$ MeV.
In the rightmost column, the r.m.s. radius for the Gogny D1 force
is also listed. 
The underline means that 
the calculated number agrees with the reference Gogny D1 result within
$10\%$.}
\label{rms-dddi}
\end{table}

\begin{figure}[htbp]
\centerline{
\includegraphics[width=8.5cm]{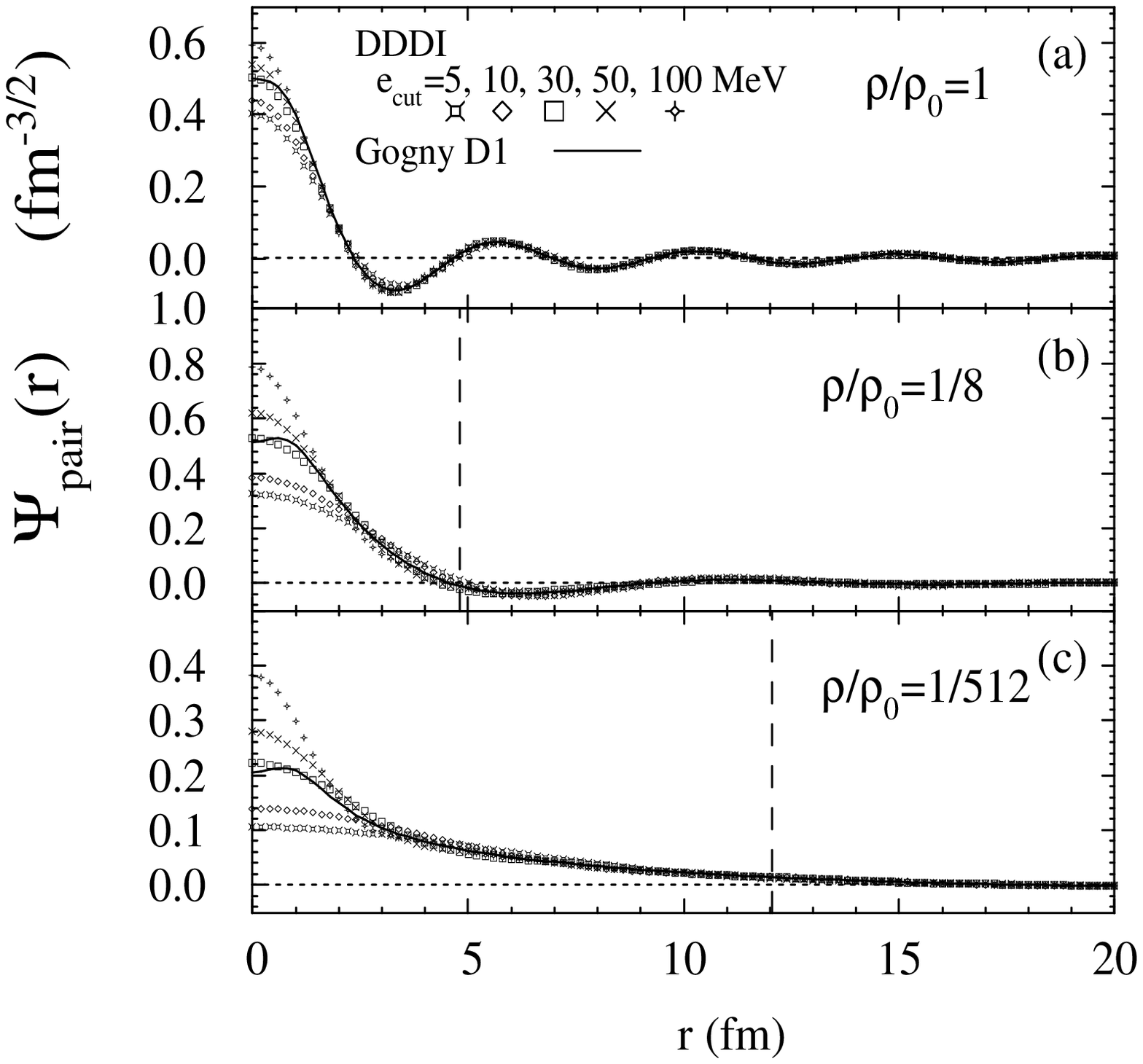}
\includegraphics[width=8.5cm]{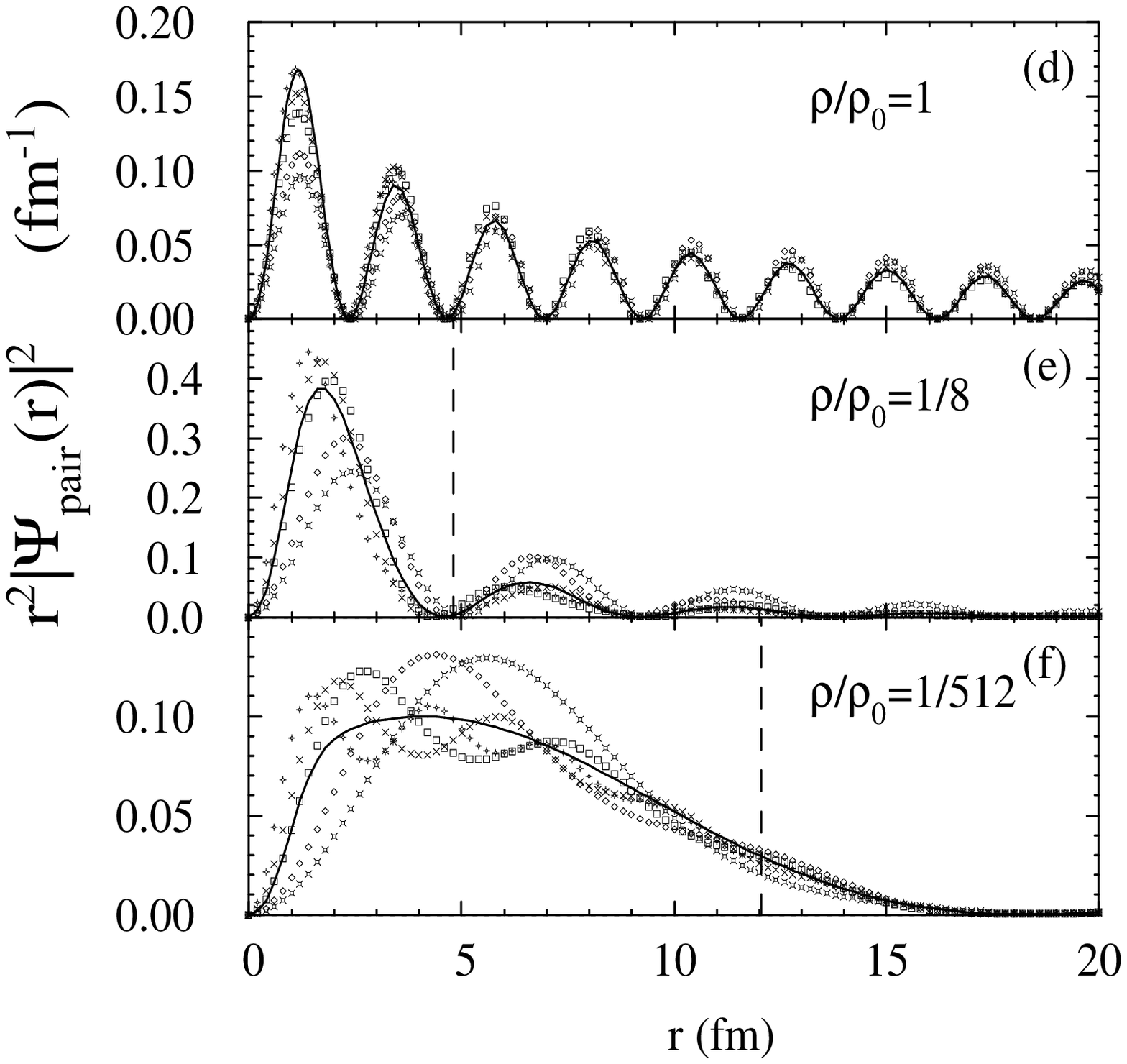}
}
\caption{
(a-c) The neutron Cooper pair wave function $\Psi_{pair}(r)$
in symmetric nuclear matter calculated with the DDDI
having different cut-off energies
$e_{cut}=5,10,30,50,100$ MeV.
 The result obtained with the Gogny D1 interaction is shown also by
the solid curve. 
(d-f) The same as (a-c) but for the probability density $r^2|\Psi_{pair}(r)|^2$.
\label{Cooper-wf-dddi}}
\end{figure}

Table \ref{rms-dddi} shows the r.m.s. radius 
$\xi_{rms}$ of the neutron Cooper pair calculated with the DDDI 
for different values of the cut-off energy.
Here $\xi_{rms}$ is calculated by using the 
Cooper pair wave function in the coordinate representation evaluated up to
$r=500$ fm. 
It is seen from Table \ref{rms-dddi} that the result apparently 
depends on the cut-off energy $e_{cut}$.  
In the cases of $\rho/\rho_0=1/64 - 1/2$, the
dependence of $\xi_{rms}$ on $e_{cut}$ is very strong. The values 
calculated with the small cut-off energy $e_{cut}=5, 10$ MeV 
largely deviate from
those obtained with the Gogny force even though the interaction strength
is chosen to reproduce the same reference pairing gap.
We consider that
the small cut-off energies $e_{cut}=5, 10$ MeV are unacceptable
since they fail to describe the small size $\xi_{rms} \sim 5$ fm
of the neutron Cooper pair at the density around  
$\rho/\rho_0=10^{-2} - 0.1$. If we require that the DDDI reproduces
 the r.m.s. radius of the neutron
Cooper pair within an accuracy of 10\% 
in the whole density region of interest, 
use of a large value of the cut-off energy satisfying $e_{cut}\gesim 50$ MeV 
is suggested.

To see roles of the cut-off energy in more details,
we show in Fig.\ref{Cooper-wf-dddi} 
the neutron Cooper pair wave functions $\Psi_{pair}(r)$ obtained
for $e_{cut}=5,10,30,50,100$ MeV.
The plot of $\Psi_{pair}(r)$ indicates clearly 
that the Cooper pair wave function depends sensitively on the 
cut-off energy $e_{cut}$. 
If we adopt the small cut-off energies $e_{cut}=5,10$ MeV
the wave function obtained with the DDDI  
 fails to produce the strong spatial correlation
at the short relative distances $r \lesim 3$ fm, 
which is the characteristic feature 
of the neutron
Cooper pair wave function common to the Gogny and G3RS forces.
(Fig.\ref{Cooper-wf-dddi} shows only the Gogny result for comparison,
but we remind the reader of Fig.\ref{Cooper-wf} where the G3RS case
is also shown.)
The plot of the probability density
$r^2|\Psi_{pair}(r)|^2$ at the density $\rho/\rho_0=1/8$ 
indicates that even the wave function at
larger distances is not described well if the small cut-off energies
$e_{cut}=5, 10$ MeV are adopted. This is nothing but the 
difficulty mentioned above in describing the r.m.s. radius with these
small cut-off energies.
If we use a large cut-off energy, say $e_{cut} \gesim 30-50$ MeV,
the wave function at large distances converges reasonably to that
obtained with the Gogny force. Concerning the wave function at short relative
distances $r \lesim 3$ fm, on the other hand, we find no convergence
with respect to the cut-off energy. The
value of the wave function $\Psi_{pair}(0)$ at zero relative distance
$r=0$ increases monotonically with increasing $e_{cut}$. (Increasing further $e_{cut}\rightarrow \infty$,  
$\Psi_{pair}(r)$ will approach to the one for the 
regularized delta interaction model shown in Fig.\ref{Cooper-wf-delta}, and 
the value $\Psi_{pair}(0)$ at $r=0$ will diverge.)

It may be possible to regard $e_{cut}$ as a parameter which simulates 
the finite range of the neutron-neutron interaction.
It is then reasonable to require that 
the wave function $\Psi_{pair}(r)$ of the DDDI model with an
appropriate choice of $e_{cut}$ describes
that of the Gogny force at distances $r\lesim 3$ fm (within 
the interaction range) as well as at larger distances. 
In the case of $\rho/\rho_0=1/8$, for example, 
this requirement is approximately satisfied if we choose
$e_{cut}=30$ or $50$ MeV, see Fig.\ref{Cooper-wf-dddi}(b).
At $\rho/\rho_0=1/512$, a good description of the wave function is
obtained with  $e_{cut}=30$ MeV
(Fig.\ref{Cooper-wf-dddi}(c)),  and similarly we find $e_{cut}\sim 70$ MeV for the
normal density $\rho/\rho_0=1$ (Fig.\ref{Cooper-wf-dddi}(a)). 
If we do not include in the comparison 
the wave function at very short distances
$r \lesim 1$ fm where the repulsion due to the core influences in the
case of the bare force, the constraint on the cut-off energy may be slightly
relaxed. For example, 
at the density $\rho/\rho_0=1/512$, 
the wave functions for $e_{cut}=30$ and $50$ MeV
differ only by about $\lesim 20\%$ at distances
$ 1<r<3$ fm, and hence the
cut-off energy $e_{cut}=50$ MeV may also be accepted.
Within this tolerance we can choose a
value around $e_{cut} \sim 50$ MeV as the cut-off
energy which can be used commonly
in the whole density region of interest.
This value can be compromised with the constraint $e_{cut} \gesim 50$ MeV
which we obtained from the condition on the r.m.s. radius of the neutron
Cooper pair.

It is interesting to note that 
cut-off quasiparticle energies around $E_{cut}=50-70$ MeV have been
employed in many of recent HFB applications to finite nuclei
\cite{DobHFB,DobHFB2,DD-Dob,DD-mix,Matsuo05,YuBulgac,Grasso,Stoitsov03,Teran,Bender}.
These cut-off energies are consistent with the 
constraint $e_{cut} \sim 50$ MeV 
suggested from the above analysis. 
Much smaller cut-off energies $\lesim 10$ MeV adopted in early applications
of the DDDI \cite{DDpair-Taj,DDpair-Tera} are not appropriate from the view point of
the spatial structure of the neutron Cooper pair wave function.
In Ref.\cite{Bertsch91} the cut-off energy of $20$ MeV
for the single-particle energy (40 MeV in the center of mass
frame energy) was shown to describe reasonably 
the scattering wave function at zero energy.
This cut-off energy is not very different from the cut-off energy $e_{cut} \sim 30$ MeV
which we find most reasonable (among the selected examples) in the lowest density case
$\rho/\rho_0=1/512$.
In the the delta interaction model adopted in Ref.\cite{Esbensen97} 
the cut-off energy is examined with respect to 
the low-energy scattering 
phase shift in the $^1S$ channel. The cut-off value adopted
is around 5-10 MeV in the single-particle energy 
($9-20$ MeV for the center of mass frame energy), 
in disagreement with our value $e_{cut} \sim 30$ MeV.
The difference seems to 
originate from different strategies to the delta interaction: 
the momentum dependence of the interaction matrix element
is accounted for by the cut-off energy in Ref.\cite{Esbensen97}
while it is taken into account in the present approach mostly 
through the density dependent interaction strength $V_0[\rho]$.

\subsection{DDDI parameters}\label{parm-sec}

It is useful to parameterize the interaction strength of the DDDI
in terms of a
simple function of the density. The following form is often assumed
\cite{Bertsch91,Garrido,DobHFB2,DD-Dob,DD-mix,DDpair-Taj,DDpair-Tera}:
\begin{eqnarray}\label{dddi-int-eq}
V_0[\rho]&=&v_0 \left(1 - \eta \left({\rho_{tot} \over \rho_c}\right)^\alpha\right), \\
\rho_c&=&0.16\ {\rm fm}^{-3}
\end{eqnarray}
where $\rho_{tot}$ is the total nucleon density.   
The parameters $v_0,\eta$ and $\alpha$ need to be determined.
In the works 
by Bertsch and Esbensen\cite{Bertsch91} and Garrido {\it et al.}\cite{Garrido}
the parameters are determined so that the parameterized DDDI reproduces 
the pairing gap obtained with the Gogny force 
in symmetric nuclear matter as well as the experimental s-wave scattering 
length at zero density.
We shall follow a similar line, but we add the important constraint 
that the spatial structure of the neutron Cooper pair is also
reasonably reproduced. As discussed in the subsection just above, 
this can be achieved if we constrain the cut-off energy to a value around
$e_{cut} \sim 50$ MeV. Note also 
our definition of the cut-off energy is different from that
in Refs.\cite{Bertsch91,Garrido},
as mentioned in Subsection \ref{cutoff-sec}. 

Our procedure is as follows. We consider symmetric nuclear
matter. The cut-off energy is fixed
to $e_{cut}=50$ MeV or 60 MeV for the reasons mentioned above.
We then fix the interaction strength $V_0[0]=v_0$ at zero density 
to a value $v_0$ which reproduces the scattering length $a$ in the
free space. 
$v_0$ satisfying this condition is given by\cite{Bertsch91,Garrido} 
\begin{eqnarray}\label{dddi-zero-eq}
v_0&=& -{ 2\pi^2\hbar^2 m^{-1} \over k_c - \pi/2a},\\
k_c &=& \sqrt{2me_{cut}}/\hbar.
\end{eqnarray}
If we use as the scattering length $a$ in Eq.(\ref{dddi-zero-eq})
the one associated with the Gogny force,
the pairing gap of the DDDI in the low density 
limit $\rho/\rho_0 \rightarrow 0$
coincides with that of the Gogny force. However, since the
scattering length $a=-13.5$ fm of the Gogny D1 is slightly off
the experimental value, we adopt the experimental one $a=-18.5$ fm
for Eq.(\ref{dddi-zero-eq}).  
This is equivalent to constrain the DDDI at the low density limit
by the bare nucleon force.

To determine the other parameters $\eta$ and $\alpha$ controlling
the density-dependence of the interaction strength,
we first calculate at several representative points of density 
the values of $V_0$ with which the neutron gap $\Delta_F$ 
of the Gogny D1 force is reproduced. 
We then search the parameters $\eta$ and $\alpha$ 
so that the simple function Eq.(\ref{dddi-int-eq}) fits well to the values of 
$V_0$ thus determined. 
We consider 
the density interval 
$\rho/\rho_0 \sim 10^{-2}- 1$ ($k_F \sim 0.3 - 1.4$ fm$^{-1}$)
in this fitting.
The obtained values of the parameters (denoted DDDI-D1) are shown in
Table \ref{dddi-parm}, where we list also the parameter set obtained
when we use $e_{cut}=60$ MeV instead of 50 MeV. Another
set of the parameters derived in the same way from the Gogny D1S force
(DDDI-D1S) is listed. We performed the same procedure also for the 
G3RS force (DDDI-G3RS).

\begin{figure}[htbp]
\centerline{
\includegraphics[angle=270,width=8.5cm]{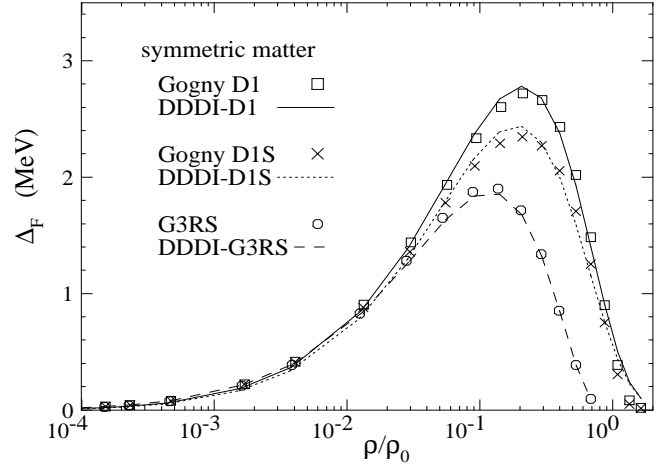}
}
\caption{The pairing gap in symmetric nuclear matter
obtained with the DDDI parameter sets shown in Table \ref{dddi-parm}
with the cut-off energy $e_{cut}=50$ MeV. The solid,
dotted and dashed curves represent the result for the
parameter sets 
DDDI-D1, DDDI-D1S, and DDDI-G3RS, respectively.
The symbols represent the gap for the reference
calculations with the Gogny D1 (square) and D1S (cross) 
forces, and the G3RS force (circle).  
\label{dddi-gap}}
\end{figure}

The pairing gap obtained with these parameterizations of
the DDDI are shown in Fig.\ref{dddi-gap}.
The resultant gap $\Delta$ agrees with that of the corresponding 
reference gap to the accuracy of about one hundred keV 
for the whole density region below $\rho/\rho_0=1$. Although
the results for $e_{cut}=60$ MeV are not shown here,  
the agreement with the reference gaps is as good as in the $e_{cut}=50$ MeV 
case. 

\begin{table}[htbp]
\begin{center}
\begin{tabular}{lcccc}
\hline
\hline
  & \hspace{5mm} $v_0$ [MeV fm$^{-3}$] \hspace{5mm}&  $\eta$  & $\alpha$   \\    
\hline
DDDI-D1\hspace{20mm} &&&& \\
\hspace{5mm}$e_{cut}=50$ MeV    & -499.9  &  0.627 & 0.55   \\ 
\hspace{5mm}$e_{cut}=60$ MeV    & -458.4  &  0.603 & 0.58   \\ 
DDDI-D1S \hspace{20mm}&&&& \\
\hspace{5mm}$e_{cut}=50$ MeV   & -499.9  &  0.652 & 0.56    \\ 
\hspace{5mm}$e_{cut}=60$ MeV   & -458.4  &  0.630 & 0.60    \\ 
DDDI-G3RS \hspace{20mm}&&&& \\
\hspace{5mm}$e_{cut}=50$ MeV   & -499.9  &  0.872 & 0.58    \\ 
\hspace{5mm}$e_{cut}=60$ MeV   & -458.4  &  0.845 & 0.59    \\ 
\hline
\hline
\end{tabular}
\end{center}
\caption{
The parameter sets of the density-dependent delta interaction
with the cut-off energies $e_{cut}=50$ and 60 MeV, derived from 
the procedure applied to the Gogny D1 and D1S, and the G3RS forces.
See text for details. }
\label{dddi-parm}
\end{table}

\begin{figure}[htbp]
\centerline{
\includegraphics[angle=270,width=8.5cm]{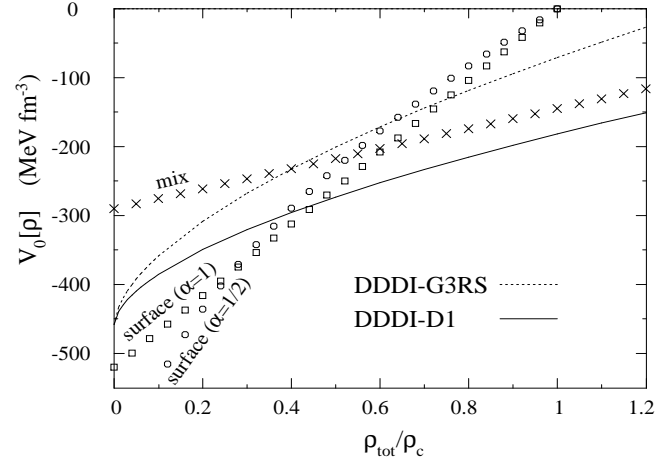}
}
\caption{The density-dependent interaction strength
$V_0[\rho]$ of the DDDI for the parameter sets
DDDI-D1 (solid curve) and DDDI-G3RS (dotted curve) with
$e_{cut}=60$ MeV. For comparison, $V_0[\rho]$ 
for the phenomenological DDDI parameters of the surface and mixed types are also
shown by the symbols. See the text for details.
\label{dddi-v0}}
\end{figure}

It is noticed in Table \ref{dddi-parm} that the parameter $\alpha$ 
takes a similar value $\alpha = 0.58-0.60$ (in the case of $e_{cut}=60$ MeV)
for all of DDDI-D1, DDDI-D1S, and DDDI-G3RS, 
while the difference between the Gogny forces (DDDI-D1,-D1S) and 
the G3RS force (DDDI-G3RS) is readily recognized 
in the value of $\eta$. It is seen also that the difference
in the cut-off energy influences slightly the values of $v_0$, 
$\eta$ and $\alpha$. If we compare our result with
that of Ref.\cite{Garrido},
our parameter values $\eta=0.60-0.63$ for the prefactor
and $\alpha=0.55-0.58$ for the power imply
stronger density dependence in $V_0[\rho]$ than that in 
Ref.\cite{Garrido}, where
the
parameters are determined as $\eta=0.45, \alpha=0.47$ from a
similar fitting to the gap with the Gogny D1 force.
This is due to the difference in the cut-off schemes mentioned
in Subsection \ref{cutoff-sec}.
Since the chemical
potential $\mu$ increases with the density, the 
energy window measured from the chemical potential $\mu$
decreases with increasing the density in the scheme of Ref.\cite{Garrido}
where the cut-off $e(k) < 60$MeV is adopted for all density
while in our cut-off
scheme the energy window
is kept constant $e_{cut}=50,60$ MeV independent of the density.
Consequently the stronger density dependence in $V_0[\rho]$
is needed in our case.

It may be interesting to compare our DDDI parameter sets with those
determined phenomenologically from the experimental 
pairing gap in the ground states of finite nuclei. Such a comparison 
is made in
Fig.\ref{dddi-v0}, where the density dependent interaction strength
$V_0[\rho]$ is plotted. The phenomenological DDDI's employed here are the
so called surface and mixed types, for which the prefactor parameter 
$\eta$ is fixed to $\eta=1$ and $\eta=0.5$, respectively. 
The power parameter
is assumed as $\alpha=1$ for the mixed type\cite{DD-mix}, 
while for the surface type 
we choose here $\alpha=1$ or 1/2. 
(Note that the power parameter in the surface 
type DDDI was investigated in Ref.\cite{DD-Dob}, 
and $ 1/2 \lesim \alpha \lesim 1$
is suggested as a  reasonable range of the parameter.)
The strength of the surface type DDDI used here is  
$v_0=-521$ MeV fm$^{-3}$ for $\alpha=1$, and
$v_0=-781$ for $\alpha=1/2$, 
taken from Ref.\cite{DD-Dob}, 
where the value of $v_0$ is determined by a Skyrme HFB calculation for
$^{120}$Sn  to reproduce
the gap $1.25$ MeV. 
The equivalent energy cut-off 60 MeV adopted in the HFB calculation 
corresponds to our cut-off $e_{cut}=60$ MeV.
In the case of the mixed type DDDI, the adopted strength is
$v_0=-290$ MeV fm$^{-3}$ derived from
the same condition on $^{120}$Sn.
It is seen in Fig.\ref{dddi-v0} that the density dependence in 
the present parameterizations of $V_0[\rho]$ is mild in the range
$\rho_{tot}/\rho_c  \gesim 0.4$, resembling more 
to that of the mixed
type DDDI than to that of the surface type.  
At lower density $\rho_{tot}/\rho_c \lesim 0.3$ the density dependence  
is stronger
than that of the mixed type DDDI. Note that 
the interaction strength $V_0[\rho]$ itself is not large: it takes
the values between those of the mixed and the surface
type DDDI's in this low density region. The above comparisons suggest that
the density dependence in 
the present parameterization 
of $V_0[\rho]$ may not be very unrealistic for applications to 
finite nuclei. It could be also suggested that the present parameterization
will be free from the problems pointed to in Ref.\cite{DD-Dob} for
strongly density dependent DDDI's such as 
the surface type DDDI with small values
of the power $\alpha\lesim 1/2$. 
For more definite conclusions, we need to make more quantitative analyses
using
the HFB calculation performed directly for finite nuclei. 
It is also interesting to compare
with a new approach to the DDDI with a microscopically 
derived cut-off factors\cite{Duguet}.
This is, however, beyond the scope of this paper,
and such analyses will be pursued in future.

\section{Conclusions}\label{concl-sec}

We have analyzed the spatial structure of
the neutron Cooper pair 
obtained with the BCS approximation for neutron and symmetric nuclear
matter
using a bare force and the effective Gogny interaction.  
The size of the Cooper pair varies significantly
with the density: its r.m.s. radius $\xi_{rms}$ becomes 
as small as $\sim 5$ fm around 
$\rho/\rho_0 \approx 10^{-2}-0.2$, and $\xi_{rms}$ 
smaller than the average inter-neutron distance $d$ is 
realized in a very wide range
$\rho/\rho_0\approx 10^{-4}-0.1$ at low density. 
The analysis of the Cooper pair wave function
indicates that the probability for the spin up and down neutrons in the
pair to be correlated within the
average inter-neutron distance $d$ exceeds more than 0.8 in this density
range. The strong spatial correlation
at short relative distances is also seen for modest density 
$\rho/\rho_0 \sim 0.5$, at which the concentration of the pair neutrons
within the interaction range $\sim 3$ fm reaches about 0.5. 
These observations suggest that   
the spatial di-neutron correlation is strong, at least in the level of the
mean-field approximation, in 
low-density superfluid uniform matter  
in the wide range of density $\rho/\rho_0\approx 10^{-4}-0.5$. 
The essential feature does not depend on the interactions.

We have investigated the behaviors of the strong di-neutron correlation
in connection with the crossover phenomenon
between the conventional pairing of the weak coupling BCS type and 
the Bose-Einstein condensation of the bound neutron pairs. Comparing with
the analytic BCS-BEC crossover model assuming a contact interaction, we found
that the density region 
$\rho/\rho_0\approx 10^{-4}-10^{-1}$ 
corresponds to the domain of the BCS-BEC crossover.

We have examined also how the density dependent delta interaction (DDDI)
combined with a finite cut-off energy can describe the spatial
correlation of the neutron Cooper pair. 
The spatial correlation at short relative distances and the
r.m.s. radius of the pair are described consistently in a wide density
region $0<\rho/\rho_0\lesim 1$ provided that we adopt 
a cut-off energy around $e_{cut}\sim 50$ MeV defined with respect to the 
chemical potential.  
We have derived a possible parametrization of the DDDI, which
satisfies this new condition on top of the constraints
on the gap in symmetric nuclear matter and on the scattering length
in the free space.
The new DDDI parameterizations may be consistent with or at least not strongly
contradictory to the phenomenological DDDI's derived from the gap 
in finite nuclei.

\section*{Acknowledgments}

The author thanks P.~Schuck and M.~Hjorth-Jensen for useful discussions.
He thanks  also the Yukawa Institute for Theoretical Physics at
Kyoto University and the Institute for Nuclear Theory at University
of Washington. Discussions during the YITP workshop YITP-W-05-01 on
``New Developments in Nuclear Self-Consistent Mean-Field Theories''
and the INT workshop ``Towards a universal density functional
for nuclei'' in the program INT-05-3 ``Nuclear Structure Near the Limits
of Stability'' were useful to complete this work.
Discussions with the members of the
Japan-U.S. Cooperative Science Program
``Mean-Field Approach to Collective Excitations in Unstable
Medium-Mass and Heavy Nuclei'' are also acknowledged.
This work was supported by the Grant-in-Aid for Scientific
Research (No. 17540244) from the Japan Society for the Promotion
of Science.


\begin{thebibliography}{99}


\bibitem{BM2}
A.~Bohr and B.~R.~Mottelson, {\it Nuclear
Structure} vol. II  (Benjamin, 1975).


\bibitem{Shimizu89}
Y.~R.~Shimizu, J.~D.~Garrett, R.~A.~Broglia, M.~Gallardo, and
E.~Vigezzi, 
Rev. Mod. Phys. {\bf 61}, 131 (1989).


\bibitem{TT93}
T.~Takatsuka and R.~Tamagaki, Prog. Theor. Phys. Suppl. No. 112, 27 (1993).




\bibitem{Lombardo-Schulze}
U.~Lombardo and H.-J. Schulze,
Lecture Notes in Physics (Springer 2001), Vol.578, p.30.

\bibitem{Dean03}
D.~J.~Dean and M.~Hjorth-Jensen,
Rev. Mod. Phys. {\bf 75}, 607 (2003).

\bibitem{Nstar1}
H.~Heiselberg and V.~Pandharipande, 
Ann. Rev. Nucl. Part. Sci, {\bf 50}, 481 (2000).

\bibitem{Nstar2}
D.~G.~Yakovlev and C.~J.~Pethick,
Ann. Rev. Astron. Astrophys. {\bf 42}, 169 (2004).


\bibitem{Bertsch91}
G.~F.~Bertsch and  H.~Esbensen, Ann. Phys. {\bf 209}, 327 (1991). 


\bibitem{DobHFB2}
J.~Dobaczewski, W.~Nazarewicz, T.~R.~Werner, J.~F.~Berger, C.~R.~Chinn, 
and J.~Decharg\'e, Phys. Rev. C {\bf 53}, 2809 (1996).

\bibitem{DD-Dob}  
J.~Dobaczewski, W.~Nazarewicz, and P.-G.~Reinhard,
Nucl. Phys. {\bf A693}, 361 (2001). 


\bibitem{DD-mix}  
J.~Dobaczewski and W.~Nazarewicz, 
Prog. Theor. Phys. Suppl. {\bf 146}, 70 (2002).

J.~Dobaczewski, W.~Nazarewicz, and M.~V.~Stoitsov, 
Euro. Phys. J. {\bf A15}, 21 (2002).

\bibitem{Tanihata}
I.~Tanihata, 
H.~Hamagaki, O.~Hashimoto, Y.~Shida, N.~Yoshikawa,
K.~Sugimoto, O.~Yamakawa, T.~Kobayashi, 
N.~Takahashi,
Phys. Rev. Lett. {\bf 55}, 2676 (1985).


\bibitem{Tanihata-density}
I.~Tanihata, 
T.~Kobayashi, T.~Suzuki, K.~Yoshida, S.~Shimoura,
K.~Sugimoto, K.~Matsuta, T.~Minamisono, W.~Christe,
D.~Olson, and H. Wieman,
Phys. Lett. {\bf B287}, 307 (1992).

\bibitem{Ozawa}
A.~Ozawa, O.~Bochkarev, L.~Chulkov, D.~Cortina, H.~Geissel, 
M.~Hellstr\"{o}m, M.~Ivanov, R.~Janik, K.~Kimura, T.~Kobayashi, 
A.~A.~Korsheninnikov, G.~M\"{u}nzenberg, F.~Nickel, Y.~Ogawa, 
A.~A.~Ogloblin, M.~Pf\"{u}tzner, V.~Pribora, H.~Simon, B.~Sit\'{a}r, 
P.~Strmend, K.~S\"{u}mmerer, T.~Suzuki,
I.~Tanihata, M.~Winkler, and K.~Yoshida,
Nucl. Phys. {\bf A691}, 599 (2001).

\bibitem{Hansen}
P.~G.~Hansen and B.~Jonson, Europhys. Lett. {\bf 4}, 409 (1987).
Nucl. Phys. {\bf A632}, 383 (1998).


\bibitem{Ikeda} 
K.~Ikeda, INS Report JHP-7 (1988); Nucl. Phys. {\bf A538}, 355c (1992).

\bibitem{Zhukov}
M.~V.~Zhukov, B.~V.~Danilin, D.~V.~Fedorov, J.~M.~Bang,
I.~J.~Thompson, and J.~S.~Vaagen,
Phys. Rep. {\bf 231}, 151 (1993).

\bibitem{Barranco01}
F.~Barranco, P.~F.~Bortignon, R.~A.~Broglia, G.~Col\'{o}, and E.~Vigezzi,
Eur. Phys. J. {\bf A11}, 385 (2001).

\bibitem{Aoyama}  
S.~Aoyama, K.~Kat\={o}, and K.~Ikeda, Prog. Theor. Phys. Suppl. 
{\bf 142}, 35 (2001).

T.~Myo, S.~Aoyama, K.~Kat\={o}, and K.~Ikeda, Prog. Theor. Phys. 
{\bf 108}, 133 (2002).

\bibitem{Hagino05}
K.~Hagino and H.~Sagawa, 
Phys. Rev. C {\bf 72}, 044321 (2005).

\bibitem{Sackett}
D.~Sackett, 
K.~Ieki, A.~Galonsky, C.~A.~Bertulani, H.~Esbensen, J.~J.~Kruse, 
W.~G.~Lynch, D.~J.~Morrissey, N.~A.~Orr, B.~M.~Sherrill, H.~Schulz, 
A.~Sustich, J.~A.~Winger, F.~De\'{a}k, \'{A}.~Horv\'{a}th, and \'{A}.~Kiss,
Z.~Seres, J.~J.~Kolata, R.~E.~Warner, and D.~L.~Humphrey,
Phys. Rev. C {\bf 48}, 118 (1993).

\bibitem{Shimoura}
S.~Shimoura, T.~Nakamura, M.~Ishihara, N.~Inabe, T.~Kobayashi, 
T.~Kubo, R.~H.~Siemssen, I.~Tanihata, and Y.~Watanabe,
Phys. Lett. {\bf B348}, 29 (1995).

\bibitem{Zinser}
M.~Zinser, F.~Humbert, T.~Nilsson, W.~Schwab, H.~Simon, T.~Aumann, 
M.~J.~G.~Borge, L.~V.~Chulkov, J.~Cub, Th.~W.~Elze, H.~Emling, H.~Geissel, 
D.~Guillemaud-Mueller, P.~G.~Hansen, R.~Holzmann, H.~Irnich, B.~Jonson, 
J.~V.~Kratz, R.~Kulessa, Y.~Leifels, H.~Lenske, A.~Magel, A.~C.~Mueller, 
G.~M\"{u}nzenberg, F.~Nickel, G.~Nyman, A.~Richter, K.~Riisager, 
C.~Scheidenberger, G.~Schrieder, K.~Stelzer, J.~Stroth, A.~Surowiec, 
O.~Tengblad, E.~Wajda, and E.~Zude,
Nucl. Phys. {\bf A619}, 151 (1997).

\bibitem{Ieki}
K.~Ieki, A.~Galonsky, D.~Sackett, J.~J.~Kruse, W.~G.~Lynch,
D.~J.~Morrissey, N.~A.~Orr, B.~M.~Sherrill, J.~A.~Winger, 
F.~De\'{a}k, \'{A}.~Horv\'{a}th, \'{A}.~Kiss,
Z.~Seres, J.~J.~Kolata, R.~E.~Warner, D.~L.~Humphrey,
Phys. Rev. C {\bf 54}, 1589 (1996).


\bibitem{Matsuo05}
M.~Matsuo, K.~Mizuyama, and Y.~Serizawa,
Phys. Rev. C {\bf 71},064326 (2005).


\bibitem{Ring-Schuck}
P.~Ring and P.~Schuck, {\it The Nuclear Many-Body Problem},
(Springer-Verlag, 1980).

\bibitem{Blaizot-Ripka} 
J.~-P.~Blaizot and  G.~Ripka, {\it Quantum Theory of Finite Systems} 
(The MIT press, 1986).


\bibitem{DobHFB}
J.~Dobaczewski, H.~Flocard,  and J.~Treiner, 
Nucl. Phys. {\bf A422}, 103 (1984). 

\bibitem{Bulgac} 
A.~Bulgac, preprint FT-194-1980, nucl-th/9907088.




\bibitem{Leggett}
A.~J.~Leggett, in {\it Modern Trends in the Theory of
Condensed Matter}, Lecture Note in Physics 115,
ed. by A.~Pekalski and R.~Przystawa
(Springer-Verlag, Berlin, 1980); J. de Phys. {\bf 41}, C7-19 (1980).

\bibitem{Nozieres}
P.~Nozi\`{e}res and S.~Schmitt-Rink, J. Low Temp. Phys.
{\bf 59}, 195 (1985). 

\bibitem{BCS} 
J.~Bardeen, L.~N.~Cooper, and 
J.~R.~Schrieffer, Phys. Rev. {\bf 108}, 1175 (1957).

P.~G.~de~Gennes, {\it Superconductivity of Metals and Alloys}
(Benjamin 1966).

M.~Tinkham, {\it Introduction to Superconductivity}
(McGraw-Hill 1975).




\bibitem{Melo}
C.~A.~R.~S\'{a}~de~Melo, M.~Randeria, and J.~R.~Engelbrecht,
Phys. Rev. Lett. {\bf 71}, 3202 (1993).

\bibitem{Engelbrecht}
J.~R.~Engelbrecht, M.~Randeria, and  C.~A.~R.~S\'{a}~de~Melo,
Phys. Rev. B {\bf 55}, 15153 (1997).

\bibitem{Randeria}
M. Randeria, in {\it Bose-Einstein Condensation},
ed. by A.~Griffin, D.~Snoke, and S.~Stringari
(Cambridge Univ. Press,  Cambridge, 1995).

\bibitem{Regal}
C.~A.~Regal, M.~Greiner, and D.~S.~Jin,
Phys. Rev. Lett. {\bf 92}, 040403 (2004).

\bibitem{Alm93}
T.~Alm, B.~L.~Friman, G.~R\"{o}pke, and H.~Schulz,
Nucl. Phys. {\bf A551}, 45 (1993).

\bibitem{Stein95}
H.~Stein, A.~Schnell, T.~Alm, and G.~R\"{o}pke,
Z. Phys. A {\bf 351}, 295 (1995).

\bibitem{Baldo95}
M.~Baldo, U.~Lombardo and P.~Schuck,
Phys. Rev. C {\bf 52}, 975 (1995).

\bibitem{Lombardo01a}
U.~Lombardo, and P.~Schuck,
Phys. Rev. C {\bf 63}, 038201 (2001).

\bibitem{Lombardo01b}
U.~Lombardo, P.~Nozi\`{e}res, P.~Schuck, H.-J.~Schulze, and A.~Sedrakian,
Phys. Rev. C {\bf 64}, 064314 (2001).


\bibitem{Takatsuka72}
T.~Takatsuka, Prog. Theor. Phys. {\bf 48}, 1517 (1972).

\bibitem{Takatsuka84}
T.~Takatsuka, Prog. Theor. Phys. {\bf 71}, 1432 (1984).

\bibitem{Baldo90}
M.~Baldo, J.~Cugnon, A.~Lejeune, and U.~Lombardo, 
Nucl. Phys. {\bf A515}, 409 (1990).


\bibitem{Oslo96}
{\O}.~Elgar{\o}y, L.~Engvik, M.~Hjorth-Jensen, E.~Osnes,
Nucl. Phys. {\bf A604}, 466 (1996).

\bibitem{Khodel}
V.~A.~Khodel, V.~V.~Khodel, and J.~W.~Clark,
Nucl. Phys. {\bf 598}, 390 (1996).

\bibitem{DeBlasio97}
F.~V.~De~Blasio,  M.~Hjorth-Jensen, {\O}.~Elgar{\o}y, L.~Engvik,
G.~Lazzari, M.~Baldo, and H.-J.~Schulze,
Phys. Rev. C {\bf 56}, 2332 (1997).

\bibitem{Serra}
M.~Serra, A.~Rummel,  and P.~Ring,
Phys. Rev. C {\bf 65}, 014304 (2001).

\bibitem{Garrido01}
E.~Garrido, P.~Sarriguren, E.~Moya~de~Guerra, U.~Lombardo, P.~Schuck, and H.~J.~Schulze,
Phys. Rev. C {\bf 63}, 037304 (2001).  

\bibitem{Kucharek89}
H.~Kucharek, P.~Ring, P.~Schuck, R.~Bengtsson, and M.~Girod,
Phys. Lett. {\bf B216}, 249 (1989).

H.~Kucharek, P.~Ring, and P.~Schuck, 
Z. Phys. {\bf 119}, 119 (1989).

\bibitem{Sedrakian03}
A.~Sedrakian, T.~T.~S. Kuo, H.~M{\"u}ther, and P.~Schuck,
Phys. Lett. {\bf B576}, 68 (2003).


\bibitem{Gogny} 
J.~Decharg\'{e} and D.~Gogny, Phys. Rev. {\bf C21}, 1568 (1980).





\bibitem{DDpair-Chas} 
R.~R.~Chasman, 
Phys. Rev. C {\bf 14}, 1935 (1976).

\bibitem{DDpair-Tera} 
J.~Terasaki, P.-H.~Heenen, P.~Bonche, J.~Dobaczewski, and H.~Flocard, 
Nucl. Phys. {\bf A593}, 1 (1995).

\bibitem{DDpair-Taj}
N.~Tajima, P.~Bonche, H.~Flocard, P.-H.~Heenen, M.~S.~Weiss,
Nucl. Phys. {\bf A551}, 434 (1993).

\bibitem{Fayans96}
S.~A.~Fayans and D.~Zawischa, Phys. Lett. {\bf B383}, 19 (1996).

\bibitem{Fayans00}
S.~A.~Fayans, S.~V.~Tolokonnikov,
E.~L.~Trykov, and D.~Zawischa, Nucl. Phys. {\bf A676}, 49 (2000).


\bibitem{Garrido}
E.~Garrido, P.~Sarriguren, E.~Moya~de~Guerra, and P.~Schuck,
Phys. Rev. C {\bf 60}, 064312 (1999).  


\bibitem{Esbensen97}
H.~Esbensen, G.~F.~Bertsch, and K.~Hencken,
Phys. Rev. C {\bf 56}, 3054 (1997). 


\bibitem{Bulren}
A.~Bulgac and Yongle~Yu,
Phys. Rev. Lett. {\bf 88}, 042504 (2002).

A.~Bulgac,
Phys. Rev. C {\bf 65}, 051305(R) (2002).


\bibitem{YuBulgac}
Yongle~Yu and A.~Bulgac,
Phys. Rev. Lett. {\bf 90}, 222501 (2003).

\bibitem{Matsupre}
M.~Matsuo, K.~Mizuyama, and Y.~Serizawa,
J. of Phys. Conf. Ser. {\bf 20}, 113 (2005);
M.~Matsuo, Proc. of the YITP workshop 
"New Developments in Nuclear Self-Consistent Mean-Field Theories",
Soryushiron Kenkyu (Kyoto) {\bf 112}, B59 (2005).


\bibitem{Tamagaki}
R.~Tamagaki, Prog. Theor. Phys. {\bf 39}, 91 (1968).

\bibitem{Aexp}
G.~F.~de~T\'{e}ramond and B.~Gabioud,
Phys. Rev. C {\bf 36}, 691 (1987).


\bibitem{Gogny-D1S}
J.~F.~Berger, M.~Girod,  and  D.~Gogny,
Comp. Phys. Comm. {\bf 63}, 365 (1991).


\bibitem{Chen86}
J.~M.~C. Chen, J.~W.~Clark, E.~Krotscheck, and R.~A.~Smith,
Nucl. Phys. {\bf A451}, 509 (1986).

\bibitem{Chen93}
J.~M.~C. Chen, J.~W.~Clark, R.~D.~Dav\'{e}, and V.~V.~Khodel,
Nucl. Phys. {\bf A555}, 59 (1993).


\bibitem{Ainsworth89}
T.~L.~Ainsworth, J.~Wambach, and D.~Pines, 
Phys. Lett. {\bf B222}, 173 (1989).

\bibitem{Wambach93}
J.~Wambach, T.~L.~Ainsworth,  and D.~Pines, 
Nucl. Phys. {\bf A555}, 128 (1993).

\bibitem{Schulze96}
H.-J.~Schulze, J.~Cugnon, A.~Lejeune, M.~Baldo,  and U.~Lombardo, 
Phys. Lett. {\bf B375}, 1 (1996).


\bibitem{Schulze01}
H.-J.~Schulze, A.~Polls, and A.~Ramos,
Phys. Rev. C {\bf 63}, 044310 (2001).

\bibitem{Lombardo01}
U.~Lombardo, P.~Schuck, and W.~Zuo,
Phys. Rev. C {\bf 64}, 021301(R) (2001).

\bibitem{Shen03}
C.~Shen, U.~Lombardo, P.~Schuck, W.~Zuo, and N. Sandulescu,
Phys. Rev. C {\bf 67}, 061302(R) (2003).

\bibitem{Schwenk}
A.~Schwenk, B.~Friman, G.~E.~Brown,
Nucl. Phys. {\bf A713}, 191 (2003).

\bibitem{Lombardo04}
U.~Lombardo, P.~Schuck, and C.~Shen,
Nucl. Phys. {\bf A731}, 392 (2004).


\bibitem{Heiselberg00}
H.~Heiselberg, C.~J.~Pethick, H.~Smith, and L.~Viverit,
Phys. Rev. Lett. {\bf 85}, 2418 (2000).


\bibitem{Fabrocini05}
A.~Fabrocini, S.~Fantoni, A.~Y.~Illarionov, and K.~E.~Schmidt,
Phys. Rev. Lett. {\bf 95}, 192501 (2005).


\bibitem{Milan1}
F.~Barranco, R.~A.~Broglia, G.~Gori, E.~Vigezzi, P.~F.~Bortignon, and J.~Terasaki,
Phys. Rev. Lett. {\bf 83}, 2147 (1999).

\bibitem{Milan2}
J. Terasaki, F.~Barranco, R.~A.~Broglia, E.~Vigezzi, 
and P.~F.~Bortignon, Nucl. Phys. {\bf A697}, 127 (2002).

\bibitem{Milan3}
G.~Gori, F.~Ramponi, F.~Barranco, P.~F.~Bortignon,
R.~A.~Broglia, G.~Col\`{o} and E.~Vigezzi, 
Phys. Rev. C {\bf 72}, 011302(R) (2005).


\bibitem{phase-shift}
{\O}.~Elgar{\o}y and M.~Hjorth-Jensen, 
Phys. Rev. C{\bf 57}, 1174 (1998).



\bibitem{Marini}
M.~Marini, F.~Pistolesi, G.~C.~Strinati,
Eur. Phys. J. B {\bf 1}, 151 (1998).

\bibitem{Papenbrock}
T.~Papenbrock, and G.~F.~Bertsch,
Phys. Rev. C {\bf 59}, 2052 (1999).


\bibitem{Heiselberg01}
H.~Heiselberg,
Phys. Rev. A {\bf 63}, 043606 (2001).


\bibitem{Grasso}
M.~Grasso, N.~Sandulescu, Nguyen~Van~Giai, and R.~J.~Liotta,
Phys. Rev. C {\bf 64}, 064321 (2001).

\bibitem{Stoitsov03} 
M.~V.~Stoitsov, J.~Dobaczewski, and W.~Nazarewicz,
S.~Pittel, and D.~J.~Dean,
Phys. Rev. C {\bf 68}, 054312 (2003).

\bibitem{Teran}
E.~Ter\'{a}n, V.~E.~Oberacker, and A.~S.~Umar,
Phys. Rev. C {\bf 67}, 064314 (2003).

\bibitem{Bender}
M.~Bender, P.-H.~Heenen and P.-G.~Reinhard,
Rev. Mod. Phys. {\bf 75}, 121 (2003).
 
\bibitem{Duguet}
T.~Duguet, 
Phys. Rev. C {\bf 69}, 054317 (2004).

\end{thebibliography}
\end{document}